\documentclass[manuscript]{aastex}
\usepackage{color}
\usepackage{float}
\usepackage{graphicx}	% Including figure files
\usepackage{amsmath}	% Advanced maths commands
\usepackage{hyperref}

\def\lapp{\ifmmode\stackrel{<}{_{\sim}}\else$\stackrel{<}{_{\sim}}$\fi}
\def\gapp{\ifmmode\stackrel{>}{_{\sim}}\else$\stackrel{>}{_{\sim}}$\fi}

\newcommand{\phx}{\texttt{PHOENIX}~}                                                               

\newcommand{\tablenote}{\texttt {~~~~~~~~}}
\shorttitle{First Transients}
\shortauthors{Lifan Wang et al.}

\begin{document}

\title{A $First~ Transients$ Survey with JWST: the FLARE project}
\author{Lifan Wang, D. Baade, E. Baron, S. Bernard,   V. Bromm,  P. Brown , G. Clayton, J. Cooke,  D. Croton, C. Curtin, M. Drout, M. Doi, I. Dominguez, S. Finkelstein, A. Gal-Yam , P. Geil, A. Heger, P. Hoeflich, J. Jiang, K. Krisciunas, A. Koekemoer, R. Lunnan, K. Maeda, J. Maund, M. Modjaz, J. Mould, K. Nomoto, P. Nugent, F. Patat, F. Pacucci, M. Phillips, A. Rest, E. Regos, D. Sand, B. Sparks, J. Spyromilio, L. Staveley-Smith, N. Suntzeff, S. Uddin, B. Villarroel, J. Vinko, D. Whalen, J. Wheeler, M. Wood-Vasey, Xiaofeng Wu, Y. Yang, \& Bin Yue}
%\date{{\color {red} 
%{\LARGE 20 October version}}}

\begin{abstract} 
The {\em James~Webb~Space~Telescope} ({\em JWST}) was conceived and built to answer one of the most fundamental questions that humans can address empirically: ``How did the Universe make its first stars?''. This can be attempted in classical stare mode and by still photography - with all the pitfalls of crowding and multi-band redshifts of objects of which a spectrum was never obtained. Our First Lights At REionization (FLARE) project transforms the quest for the epoch of reionization from the static to the time domain. It targets the complementary question: ``What happened to those first stars?''.  It will be answered by observations of the most luminous events: supernovae and accretion on to black holes formed by direct collapse from the primordial gas clouds. These transients provide direct constraints on star-formation rates and the truly initial initial mass function, and they may identify possible stellar seeds of supermassive black holes.  Furthermore, our knowledge of the physics of these events at ultra-low metallicity will be much expanded. {\em JWST}'s unique capabilities will detect these most luminous and earliest cosmic messengers easily in fairly shallow observations. However, these events are very rare at the dawn of cosmic structure formation and so require large area coverage.  Time domain astronomy can be advanced to an unprecedented depth by means of a shallow field of {\em JWST} reaching 27 mag (AB) in 2~$\mu$m and 4.4~$\mu$m over a field as large as 0.1~square degree visited multiple times each year. Such a survey may set strong constraints or detect massive Population~III supernovae at redshifts beyond 10, pinpointing the redshift of the first stars, or at least their death. Based on our current knowledge of superluminous supernovae, such a survey will find one or more superluminous supernovae at redshifts above 6 in five years and possibly several direct collapse black holes. In addition, the large scale structure that is the trademark of the epoch of reionization will be detected. Although {\em JWST} is not designed as a wide field survey telescope, we show that such a wide field survey is possible with {\em JWST} and is critical in addressing several of its key scientific goals.
\end{abstract}

\maketitle
\tableofcontents

% ==============================================================================
\section{Introduction}
\label{section:Introduction}

Just as a famous early goal of the {\em Hubble~Space~Telescope} ({\em HST}) was to measure the Hubble Constant, so a well publicized early goal of the {\em James~Webb~Space~Telescope} ({\em JWST}) is to find and characterize the first stars in the Universe. Population~III (Pop~III) does not come with a handbook of how to do this. In a `shallow' survey, {\em JWST} takes us into a discovery space of transient phenomena at AB~$\sim$~27 mag of which we have no knowledge. The community needs to know about transient phenomena in the very early Universe, as transients are prompt tracers of the constituents of the Universe. This knowledge may affect the subsequent mission in significant ways. 

Supernovae (SNe) are the defining characteristic of Pop~III. When the first SNe occur, Pop~III ends, because they inject first metals into the intergalactic medium. {\it Inter alia} the early science with {\em JWST} %First Transients ERS may 
can recognize the first supernovae by their rise time. If this can be achieved, a wealth of follow up science  opens up for the community, not fully predictable at the time of launch, but possibly among the major outcomes of the mission.

In this paper we consider in turn survey science goals (\S\ref{section:TSSG}), a 0.1~square degree field survey for transients with {\em JWST} (\S\ref{section:FLARE}), and facility deliverables to allow community use of the survey (\S\ref{section:FFTD}). We draw our conclusions in \S\ref{section:Conclusion}, leaving detailed simulations for a subsequent paper.

\subsection{The Epoch of First Light}
%{\it written by by Volker Bromm}

A key gap in our understanding of cosmic history is the pre-reionization Universe, comprising the first billion years of cosmic evolution \citep{fg11,Loeb2013}. Within $\Lambda$CDM cosmology, we have an increasingly detailed theoretical picture of how the first stars and galaxies brought about the end of the cosmic dark ages, initiating the process of reionization \citep{Barkana2007,Robertson2010}, and endowing the primordial universe with the first heavy chemical elements \citep{Karlsson2013}. The initial conditions for primordial star formation are known to very high precision by probing the spectrum of primordial density fluctuations imprinted in the cosmic microwave background (CMB), carried out by {\em WMAP} and {\em Planck}. The first stars thus mark the end of precision cosmology, and their properties are closely linked to the cosmological initial conditions.

We are entering the exciting period of testing our predictions about the first stars and galaxies with a suite of upcoming, next-generation telescopes. Soon, {\em JWST} will provide unprecedented imaging sensitivity in the near- and mid-IR, to be followed by the 30--40~m class, ground-based telescopes now under development. %The Giant Magellan Telescope (GMT),
%the Thirty-Meter Telescope (TMT), and the European Extremely Large Telescope (E-ELT),
%enhanced by innovative adaptive-optics technology, will be ideally complementary
%to the {\em JWST} with their exquisite spectroscopic capabilities. 
In addition, astronomers will bring to bear facilities that can probe the cold Universe at these early times. Next to the ALMA observatory, a suite of meter-wavelength radio interferometers are being built that will probe the redshifted 21-cm radiation, emitted by the neutral hydrogen at the end of the cosmic dark ages \citep{Furlanetto2006}. The radio facilities provide a view of the raw material, the cold primordial hydrogen gas, out of which the first stars form. In addition, there are missions that aim at discovering transients originating in the violent deaths of the first stars. A prime example is the search for high-redshift gamma-ray bursts (GRBs), marking the death of progenitor stars that are massive enough to collapse into a black hole \citep{bl06a,Kumar2015}. Overall, the prospects for closing the remaining gap in understanding the entire history of the Universe, including the still elusive first billion years, are bright. This gap is actually a discontinuity in our understanding which prevents us from smoothly connecting the predictions from the earlier epoch of precision cosmology to extrapolations from the local Universe and intermediate redshifts.

\subsection{Current candidate first galaxies and Ly$\alpha$ blobs}
At the redshift of reionization ($z > 6$), emission lines from distant galaxies become increasingly difficult to observe. The brightest line, Ly$\alpha$, becomes virtually the only line accessible from the ground. Significant progress has been made in recent years  finding luminous Ly$\alpha$ emitters \citep[e.g.][]{Matthee:2015, Ouchi:2010, Sobral:2015}.

\begin{figure}[ht]
\centering
\begin{tabular}{ll}
\includegraphics[width=0.4\textwidth]{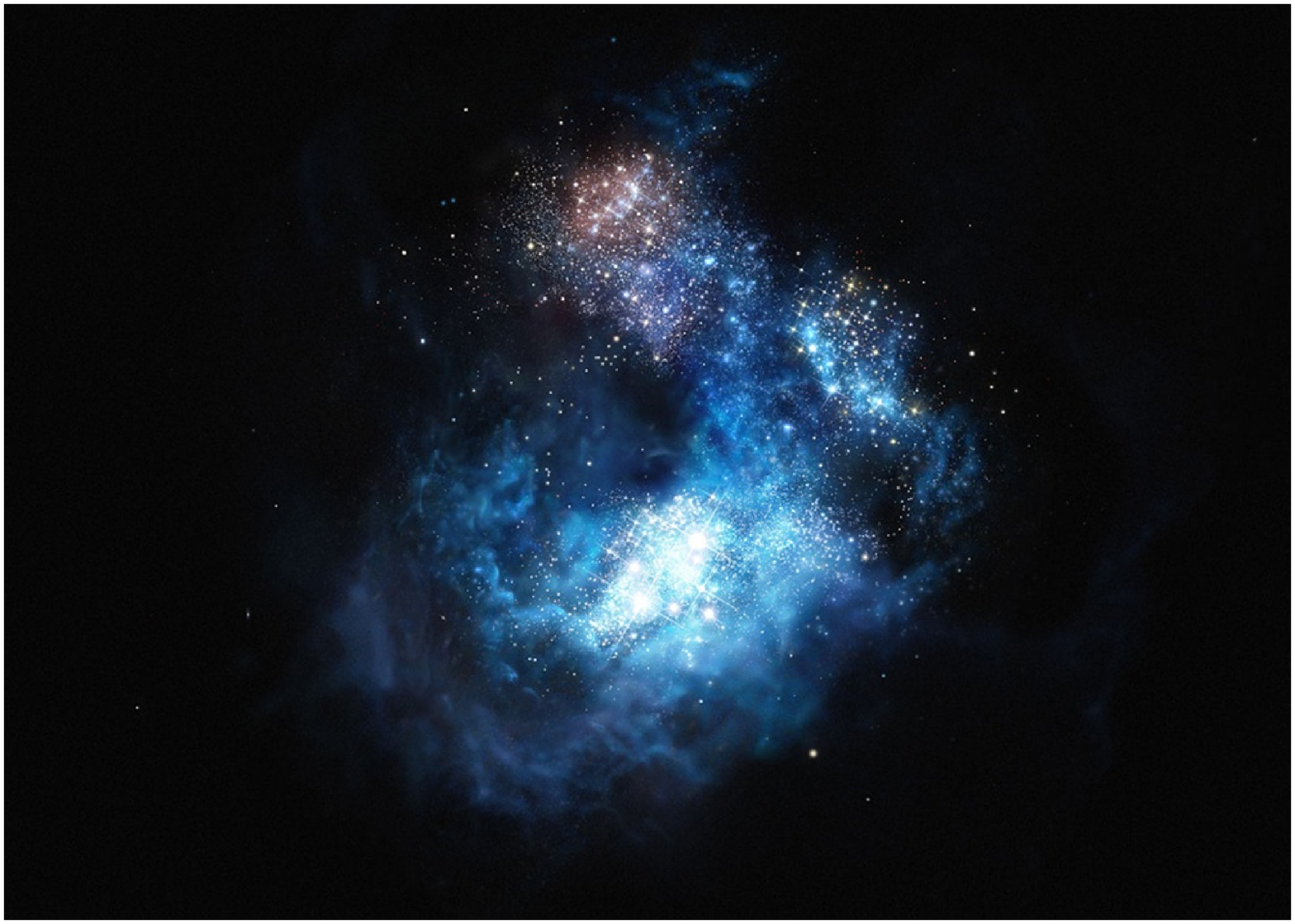} &
\includegraphics[width=0.48\textwidth]{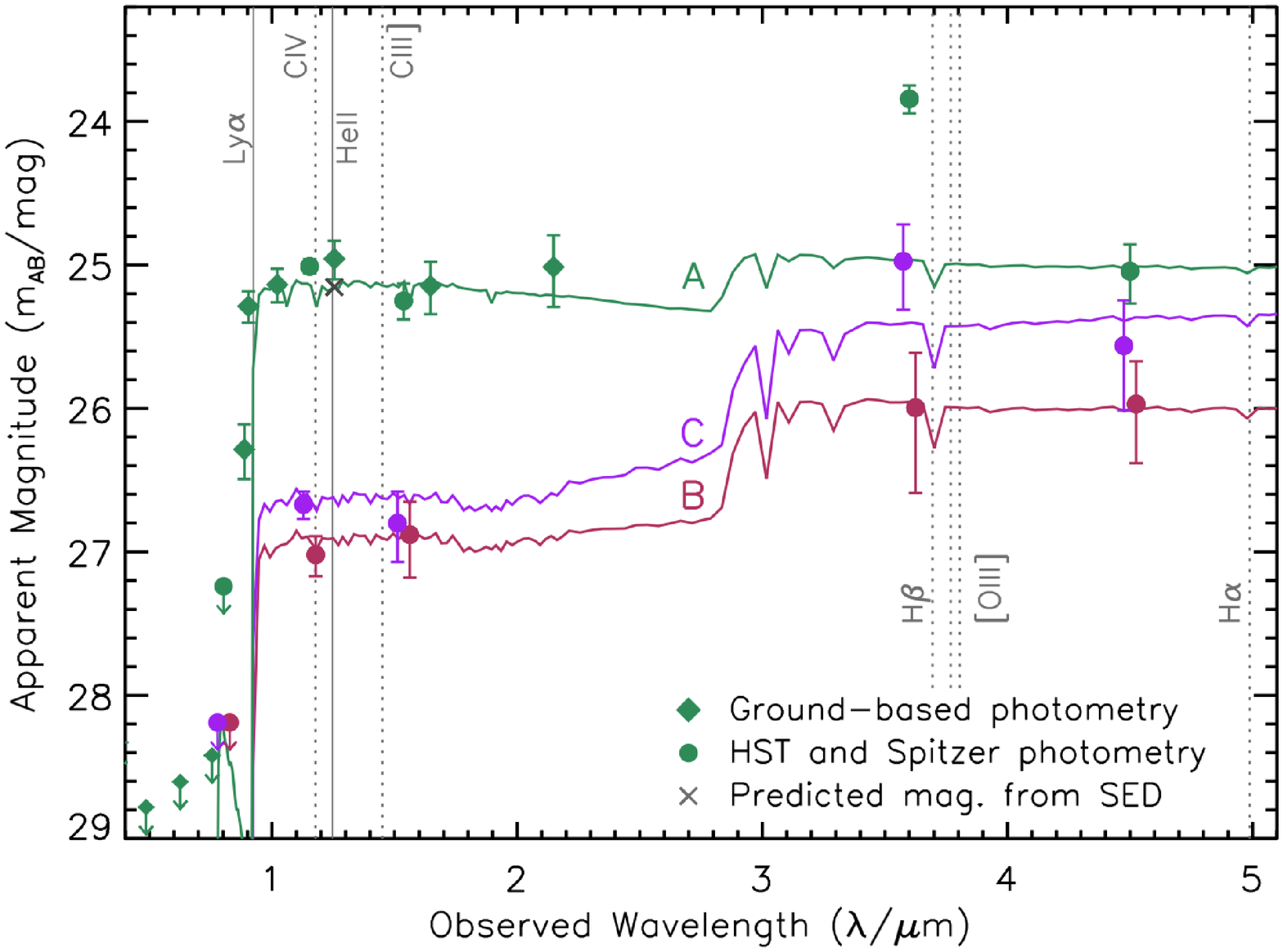} \\
\end{tabular}
\caption{Left: COSMOS Redshift 7 (abbreviated to CR7) %, shown in a big red circle, 
is a bright galaxy found in the COSMOS field \citep{Matthee:2015,Sobral:2015}. %The green circles show all Ly$\alpha$ candidates found in the COSMOS/UltraVISTA field. (Figure taken from \citet{Sobral:2015}). 
Right: The SEDs of the three components of CR7 are plotted in green (A), purple (B) and red (C), together with the best-fitting \cite{Bruzual:2003} model shown as the solid line in each case. The solid vertical grey lines show the observed wavelengths of the confirmed Ly$\alpha$ and He~II emission lines,  whereas the dotted vertical lines show for reference the wavelengths of other rest-frame UV and optical lines that have been detected or inferred in other high-redshift Lyman-break galaxies \citep{Stark:2015a,Stark:2015b,deBarros:2016,Smit:2014,Smit:2016}.}
\label{CR7}
\end{figure}

The best observational evidence on first generation stars or black hole (BH) seeds is found in a distant galaxy CR7 (Figure~\ref{CR7}). These identifications rely heavily on theoretical models of the radiation from first stars \citep[e.g.,][]{pal15,Xu:2016,Yajima:2017} or direct collapse black holes (DCBHs) (e.g. \citealt{Pacucci2017,Agarwal:2016,Agarwal:2017}; which was disputed by \citealt{Bowler:2017}).

The spectral energy distribution (SED) of CR7 is shown in Figure~\ref{CR7} \citep{Bowler:2017}. A shallow, wide-field survey with {\em JWST} is sufficiently deep to discover luminous Ly$\alpha$ emitters like CR7. \cite{Matthee:2015} found that the luminous end of the luminosity function of Ly$\alpha$ emitters at $z = 6.6$ is comparable to the luminosity function at $z = 3$--5.7, and is consistent with no evolution at the bright end since $z \sim 3$. The number density of luminous Ly$\alpha$ emitters is thus found to be much more common than expected. The space density is 1.4$^{+1.2}_{-0.7}\times10^{-5}$~Mpc$^{-3}$. From Figure~\ref{CR7}, we see that such targets can be easily discovered in a survey that goes to 27th mag in 2~$\mu$m and 4.4~$\mu$m. They are characterized by very red colors in this wavelength range. Note, however, \cite{Bowler:2017} found a blue $[3.6]-[4.5]$ color based on {\em Spitzer} data, and argued that CR7 is likely a low-mass, narrow-line active galactic nucleus (AGN) or is associated with a young, low-metallicity ($\sim 1/200$~Z$_\odot$) star-burst. 

In the design of FLARE, we expect to discover 20--50 luminous Ly$\alpha$ emitters per unit redshift bin at $z \sim 6$--10 assuming a non-evolving rate of luminous Ly$\alpha$ emitters.

%{\bf Variability of Ly$\alpha$ emitters ??? Any publications? From IUE ovservations we know Ly$\alpha$ flux varies with the continuum or broad band photometry. If this applies at high-redshift, we expect Ly$\alpha$ emitters to show variations similar to AGNs in the UV.}

\subsection{AGN Variability}

Among the high-redshift transients FLARE will detect are AGN. These are signposts to the formation of the supermassive black holes (SMBHs) that we know are catapulted from stellar seeds to $\gapp10^9$~M$_\odot$ in the first half billion years of the Universe.
%{\it written by J Mould}\\
Our comoving volume between redshift 6 and 7 is approximately one million Mpc$^3$. If every galaxy with a mass exceeding the Milky Way's has an SMBH, that volume contains 10$^4$ SMBH. A signature of these is Tidal Distortion Events (TDEs), manifesting themselves as order of magnitude flares in rise times of $(1+z)$ times 30 days. Assuming steady growth between $z = 20$ and $z = 6$, each 10$^7$~M$_\odot$ SMBH is adding 0.01~M$_\odot$ per year. If 10\% of this material is stars, the rate of TDEs is of order ten per year. What makes these high-redshift objects detectable, as is also true of superluminous supernovae (SLSNe), is that accretion or shocks make them UV bright, more like flat spectrum sources than the 10,000~K black bodies that normal photosphere-like SNe resemble. \cite{Caplar2016} find that the amplitude of AGN variation in PTF and SDSS is typically a tenth of a magnitude, but, interestingly, this goes up by a factor of 5 in the UV, which is where FLARE will observe at high redshift.
%1. Damped Random Walk model
%2. Structure function
%3. SDSS Stripe 82; PTF/iPTF
%4. Extrapolation to the UV
%5. How do we find them at high redshift?

% ==============================================================================
\section{Transient Survey Science Goals}
\label{section:TSSG}

The purpose of the survey is to find transients as witnesses of the state of the Universe when the first stars and black holes were formed. As we shall see, an appropriate search strategy is a mosaic approach with only modestly deep individual exposures. In \S\ref{section:FLARE} a 0.1~square degree field is considered to reach into this wide-field regime and place constraints on the massive Pop~III star formation rate density.

\subsection{Population~III Supernovae}
How massive can the first stars, the so-called Population~III, grow \citep{Bromm2013}? This question is important in order to determine the nature of the death of these stars, which in turn governs the feedback exerted by them on their surroundings \citep{Maeder2012}. To address it, simulations have to be pushed into the radiation hydrodynamic regime, where protostars more massive than $\sim 10$~M$_{\odot}$ emit ionizing photons, resulting in the build-up of ultra-compact H\,{\sc II} regions \citep{Hosokawa2011, Hosokawa2016, hir13, Stacy2012}. The roughly bi-conical H\,{\sc II} regions then begin to photo-evaporate the proto-stellar disk, thus choking off the gas supply for further accretion \citep{McKee2008}. Upper masses thus reached are typically in the vicinity of a few times 10~M$_{\odot}$, such that the most likely fate encountered by the first stars is a core-collapse supernova. On the other hand, the prediction is that the primordial initial mass function (IMF) was rather broad, extending both to lower and (possibly significantly) higher masses. The latter would imply that BH remnants should be quite common as well. In rarer cases pair-instability (PI) SN progenitors would give rise to hyper-energetic explosions. See also \cite{Woosley2002}, \cite{Heger2003}, \cite{Heger2002}, \cite{Kasen2011}, and references therein.

%\begin{figure}[ht]
%\includegraphics[width=0.6\textwidth]{volker.eps}
%\caption{PI~SN rates in number per year per {\em JWST} field of view above a given redshift \citep[from][]{hum12}.
%Shown is the upper limit for weak feedback (blue line), strong feedback (red line), and an intermediate case (dashed red line). The symbols denote additional estimates from the literature.}
%\end{figure}

\subsubsection{Probing Pop~III with Transients}
Transients are the targets of choice because they are the brightest individual objects, whereas not even the first galaxies are detected in the same observations. Testing  Pop~III predictions is challenging, given that Pop~III stars are short-lived, and that rapid initial enrichment hides Pop~III star formation from view, such that even the deepest {\em JWST} exposures will typically tend to see metal-enriched, Population~II, stellar systems. Therefore, a more indirect strategy is required. A promising approach is to constrain the extreme ends of the Pop~III IMF. On the high-mass end, the search for PI~SNe in upcoming {\em JWST} surveys will tell us whether primordial star formation led to masses in excess of $\sim 150$~M$_\odot$, the threshold for the onset of the pair-production instability. Individual PI~SN events are bright enough to be detected with the {\em JWST}, but the challenge will be their low surface density, such that a wide area needs to be searched \citep{hum12,pan12a,wet12b}.

\subsubsection{Finding the First Supernovae}
%{\it written by Dan Whalen}
The cosmic dark ages ended with the formation of the first stars at $z \sim 20$, or $\sim 200$~Myr after the Big Bang \citep[e.g.,][]{fsg09,fg11,dw12}. The first stars began cosmological reionization \citep{wan04,abs06,awb07}, enriched the primeval universe with the first heavy elements \citep{mbh03,ss07,wet08a,jet09b,ss13}, populated the first galaxies \citep{jlj09,pmb11,wise12,ren15}, and may be the origin of the first SMBHs \citep{wf12,jet13,Woods2017,smidt17}. In spite of their importance to the early universe, not much is known about the properties of Pop~III stars, as not even {\em JWST} or 30--40~m class telescopes such as the Giant Magellan Telescope (GMT), the Thirty-Meter Telescope (TMT), or the Extremely Large Telescope (ELT) will be able to see them \citep{rz10}. And while the initial conditions for Pop~III star formation are well understood, having been constrained by measurements of primordial density fluctuations by both {\em WMAP} and {\em Planck}, numerical simulations cannot yet determine the masses of primordial stars from first principles \citep[e.g.,][]{fsg09,fg11,dw12}. However, some of these models suggest that they may have ranged from a few tens of solar masses to $\sim 1000$~M$_{\odot}$, with a fairly flat distribution in mass (\citealt{hir13,hir15}; see also \citealt{Woosley2017}).

High-redshift SNe could directly probe the first generations of stars and their formation rates because they can be observed at great distances and, to some degree, the mass of the progenitor can be inferred from the light curve of the explosion \citep{tet13,ds13,ds14}. 1D Lagrangian stellar evolution models \citep{hw02} predict that 8--40~M$_{\odot}$ non-rotating Pop~III stars die as core-collapse (CC) SNe and that 40--90~M$_{\odot}$ stars collapse to BHs. If the stars are in rapid rotation, they can produce a GRB \citep{bl06a,mes13a} or a hypernova \citep[HN; e.g.,][]{nom10}. 10--40~M$_{\odot}$ stars can also eject shells prior to death, and the subsequent collision of the SN ejecta with the shell can produce highly luminous events that are brighter than the explosion itself % (Type IIn SNe)
\citep[pulsational PI~SNe, or PPI~SNe;][]{wbh07,chen14a}.

At $\sim 100$~M$_{\odot}$ non-rotating Pop~III stars can encounter the pair instability. At 100--140~M$_{\odot}$ the pair instability causes the ejection of multiple, massive shells instead of the complete destruction of the star \citep{Woosley2007}.
%But %collisions between 
%these shells can, like Type IIn SNe, can produce extremely bright events %in the UV 
Rotation can cause stars to explode as PI~SNe at masses as low as $\sim 85$~M$_{\odot}$ \citep[rotational PI~SNe, or RPI~SNe;][]{cw12,Yoon2012,cwc13}. Above 260~M$_{\odot}$ stars encounter photodisintegration with 140--260~M$_{\odot}$ stars exploding as highly energetic thermonuclear PI~SNe \citep{rs67,brk67,jw11,chen14c}. %,  instability and collapse to a BH. 
Finally, at much higher masses ($\gapp$ 50,000~M$_{\odot}$) some stars may die as extremely energetic thermonuclear SNe triggered by the general relativistic instability \citep{montero12,chen14b}. Such events could occur in atomically cooled halos at $z \sim 15$--20, heralding the birth of DCBHs, and thus the first quasars \citep{jet13a,wet13a,wet13b,wet12d}. 

%added by J Vinko & Craig Wheeler

SLSNe can be another critical probe of the deaths of the first stars, because those events are bright ($M < -20.5$ mag) \citep{DeCia2017,Lunnan2017} and have UV-luminous SEDs in the rest-frame. SLSNe are typically 2--3 magnitudes brighter than SNe~Ia, and 4 or more magnitudes brighter than typical CC~SNe. Observationally, there are two known types of SLSNe. Hydrogen-rich SLSN~II are thought to gain their extremely high luminosity from the collision between the expanding SN ejecta and a dense circumstellar shell ejected prior to explosion. The mechanism of the explosion itself is less clear, because it is hidden behind the dense circumstellar cloud. Hydrogen-deficient SLSNe, called SLSN~I, do not show signs for such a violent collisional interaction, thus, they are thought to be powered by the spin-down of a magnetar resulting from the core collapse of an extremely massive progenitor star \citep{Nicholl2017}.

\begin{figure}[H]%[ht]
\centering
% \begin{center}
\includegraphics[width=0.7\linewidth]{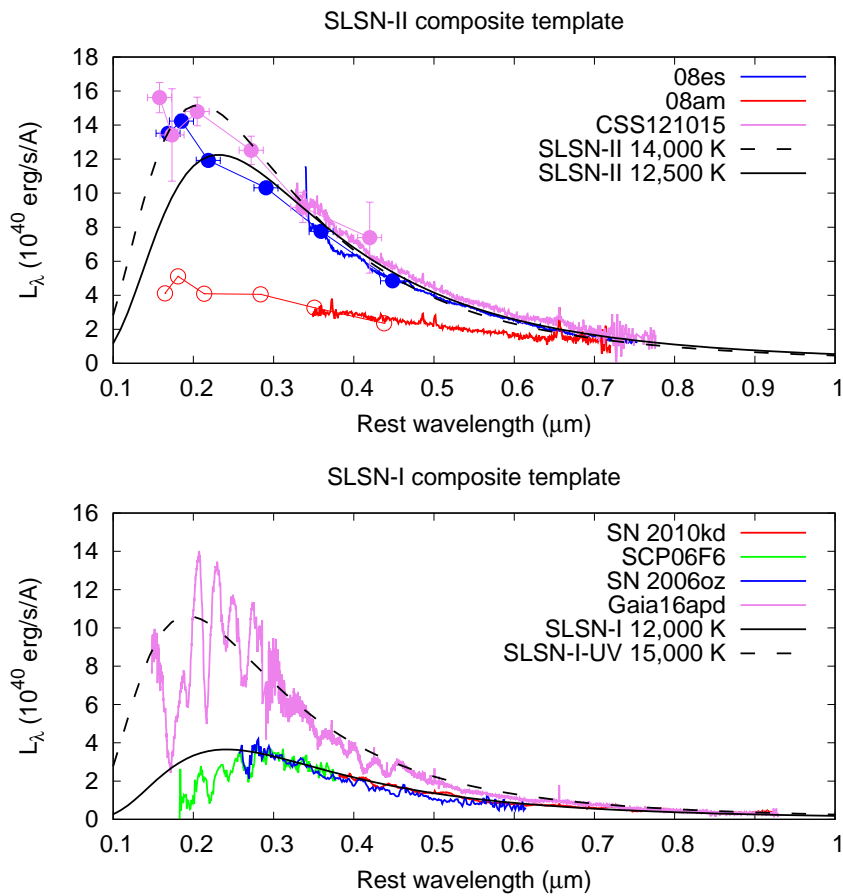}
%\includegraphics[width=0.7\linewidth]{slsneII_jwst.eps}
% \end{center}
\caption{The modeled SEDs of SLSNe at maximum light. Hydrogen-rich (SLSN-II) events are plotted in the top panel, while hydrogen-free (SLSN-I) ones are in the bottom panel. The black solid and dotted lines illustrate black body continua that fit the observations.}
\label{fig:SLSNmod}
\end{figure}

We have modeled the rest-frame SED of SLSNe~I and II using the available data of several well-observed local events (see Figure~\ref{fig:SLSNmod}) downloaded from the Weizmann Interactive Supernova Repository\footnote{\url{https://wiserep.weizmann.ac.il/}} and the Open Supernova Catalog\footnote{\url{https://sne.space}}. The UV-optical spectra taken close to maximum light, after scaling to match photometry, have been corrected for extinction and distance. Figure~\ref{fig:SLSNmod} shows the final composite spectra together with scaled black bodies that fit the continuum. Since SLSN spectra are not homogeneous, we fit two black bodies having different temperatures and scale factors (plotted with solid and dashed lines, respectively) to represent the range of luminosities for both SLSN types. 

\begin{figure}[H]%[ht]
% \begin{center}
\centering
\includegraphics[width=0.7\linewidth]{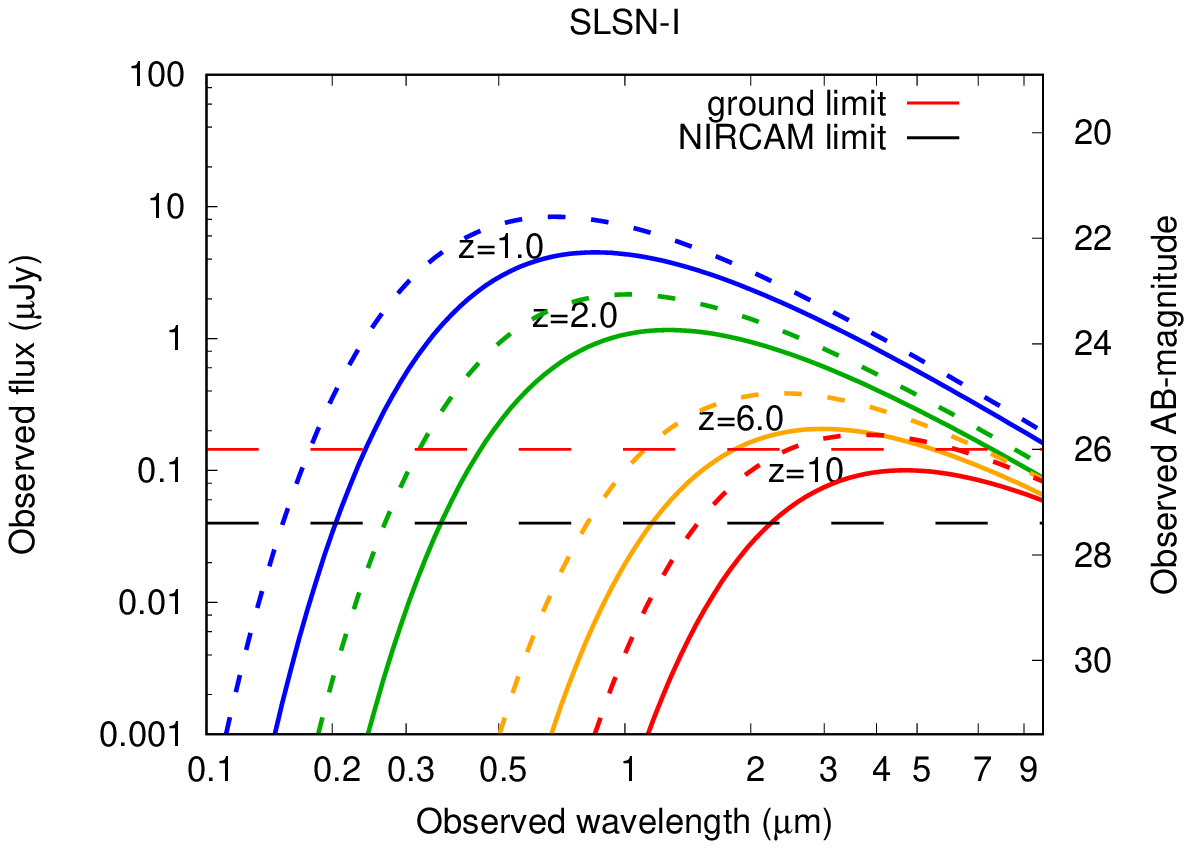}
\includegraphics[width=0.7\linewidth]{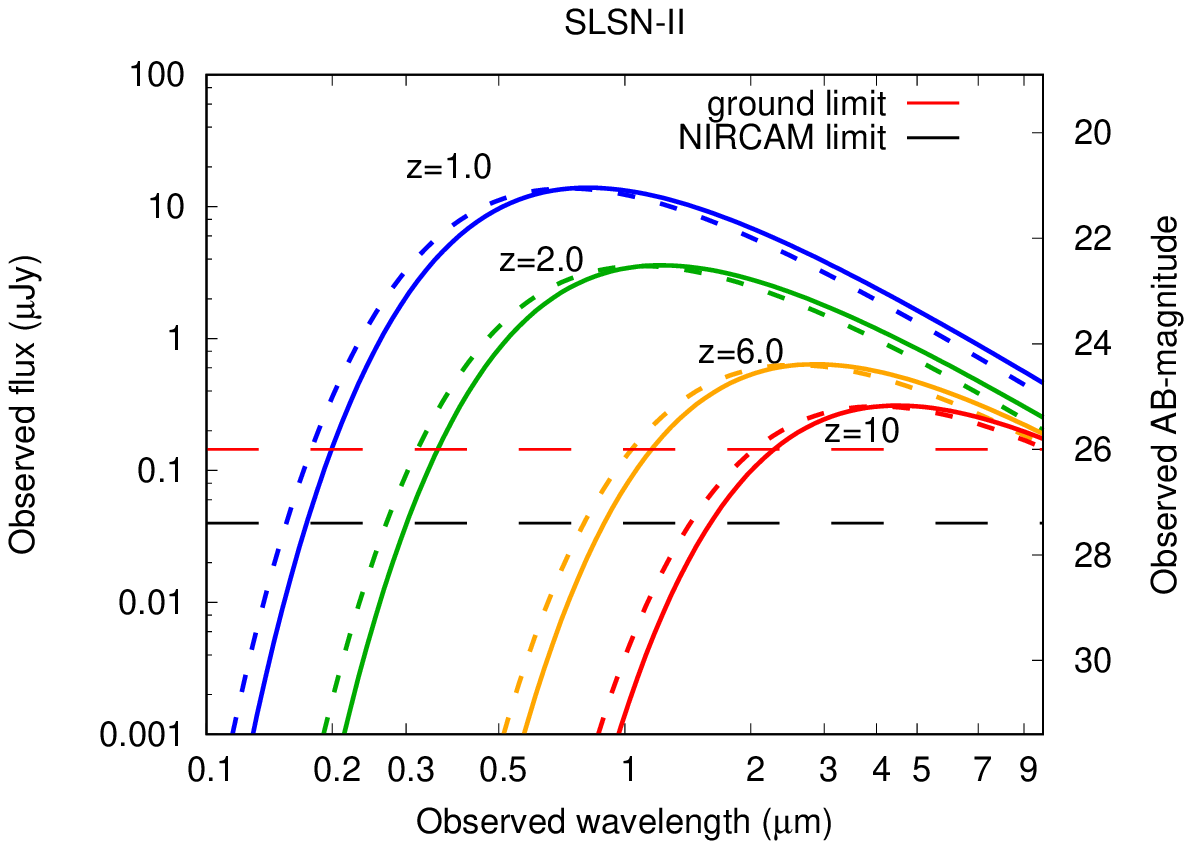}
% \end{center}
\caption{The model SEDs of SLSNe (plotted with colored solid and dashed lines) at different redshifts. Top panel: SLSN-I. Lower panel: SLSN-II. The SEDs plotted with dashed lines correspond to the models having higher temperatures (see Figure~\ref{fig:SLSNmod}). Dashed red and black horizontal lines represent the detection sensitivity limits from the ground and with {\em JWST}, respectively.}
\label{fig:SLSNz}
\end{figure}

In Figure~\ref{fig:SLSNz} we plot the model SEDs of SLSNe redshifted between $z=1$--10. It is seen that detection with {\em JWST} between 2--4~$\mu$m can be feasible for both types even at $z \sim 10$, provided SLSNe do exist at such high redshift. Figure~\ref{fig:SLSNz} suggests that the detection, in principle, could be pushed above $z = 10$, but time dilation that increases the timescale of transients by a factor of $1+z$ limits the practical discovery efficiency of the intrinsically slowly evolving SLSNe at such high redshifts.

\subsubsection{Observational constraints}
%added by J. Vinko & Craig Wheeler

]

%{\it Written by R Quimby}
The expected rate of SLSNe beyond $z>1$ can be estimated from their observed local rates. Locally, SLSNe are rare events. Relatively shallow, flux-limited surveys like CRTS, PTF, and ASASSN discover just a few SLSNe for roughly every 100 normal luminosity SNe. Obviously these surveys can search for SLSNe in much larger effective volumes than for normal luminosity SNe, so the rate of SLSNe is but a fraction of the total SN population's rate. \cite{Quimby2013} measured a SLSN-like rate of $199^{+137}_{-86}$ events/Gpc$^3$/yr (h$^3_{71}$) at $z=0.16$, which is about 1/1000th of the core-collapse rate at this redshift. \cite{Prajs2017} measure a rate that is roughly twice as high at $z=1.13$, and \cite{Cooke2012} estimate that the rate is higher by at least another factor of 5 at $2 < z < 4$. Given our current understanding that SLSNe originate with the deaths of very massive stars typically in low-metallicity environments, such an increase in rate with redshift is to be expected. The precise redshift evolution is of great value to measure as it can either be used to reveal how environmental factors affect the production of SLSNe (and thus help reveal their physical origin), or it can be used to probe how the production of the most massive stars evolves in relation to lower-mass stars \citep[i.e. the SLSN rate evolution may help reveal any changes in the stellar IMF with redshift;][]{tet13}.

%\begin{figure}[H]%[ht]
%\centering
%\includegraphics[angle=-0,width=0.6\textwidth]{tmy.eps}
%\vspace{-2 truemm}
%\caption{Rates from \cite{tet13} imply 0.3 SLSNe for z$>$6 in the ERS field. }
%\end{figure}

\subsubsection{Modelled high-redshift SLSNe rates}
%{\it { written by Paul Geil}}
%Mould et al DECamERON saw >1 per sq deg in the redshift range 5.5-6.5

Observations show a strong preference for SLSNe toward high star-forming, low-metallicity environments \citep{Lunnan2014,Leloudas2015}. As such, SLSNe are expected to trace the cosmic star formation history. This is evident from the increase in the observed volumetric rate out to $z \sim 4$ \cite[cosmic star formation peaks at $z \sim 2$--3, see, e.g.,][]{Hopkins2006,Moster2017}. The metallicity dependence is likely related to the SLSN production mechanism and dependent on progenitor properties.

To model the comoving SLSN rate as a function of redshift, we combine a star formation rate model ($\dot{\rho_*}$) with a metallicity-dependent efficiency $\epsilon_Z$ for their formation, i.e.
\begin{equation}
\dot{n}_{\rm SLSNe} (z) = \epsilon_{Z}(z) \dot{\rho}_*(z).
\label{eq:volrate}
\end{equation}
This method is based on the rate modelling of GRBs by \cite{Trenti2013}. We use DRAGONS' semi-analytical galaxy-formation model \citep{DRAGONS3}, which shows good agreement with both observations and other models. The efficiency function is the key to accounting for different SLSN progenitor models via metallicity. As progenitor models are poorly constrained, we employ the following simple, empirically-motivated prescription. Using the mean stellar metallicity of every galaxy at each simulated redshift, we calculate a SLSN production efficiency factor \citep[using stellar evolution simulations by][]{Yoon2006} for each galaxy and then average over all galaxies at that redshift. The basic form for the efficiency factor is higher efficiency for lower metallicities with an adjustable lower threshold plateau in place at higher metallicities (as SLSNe may still occur in these environments).

\begin{figure}[H]%[ht]
\centering
\includegraphics[width=12cm]{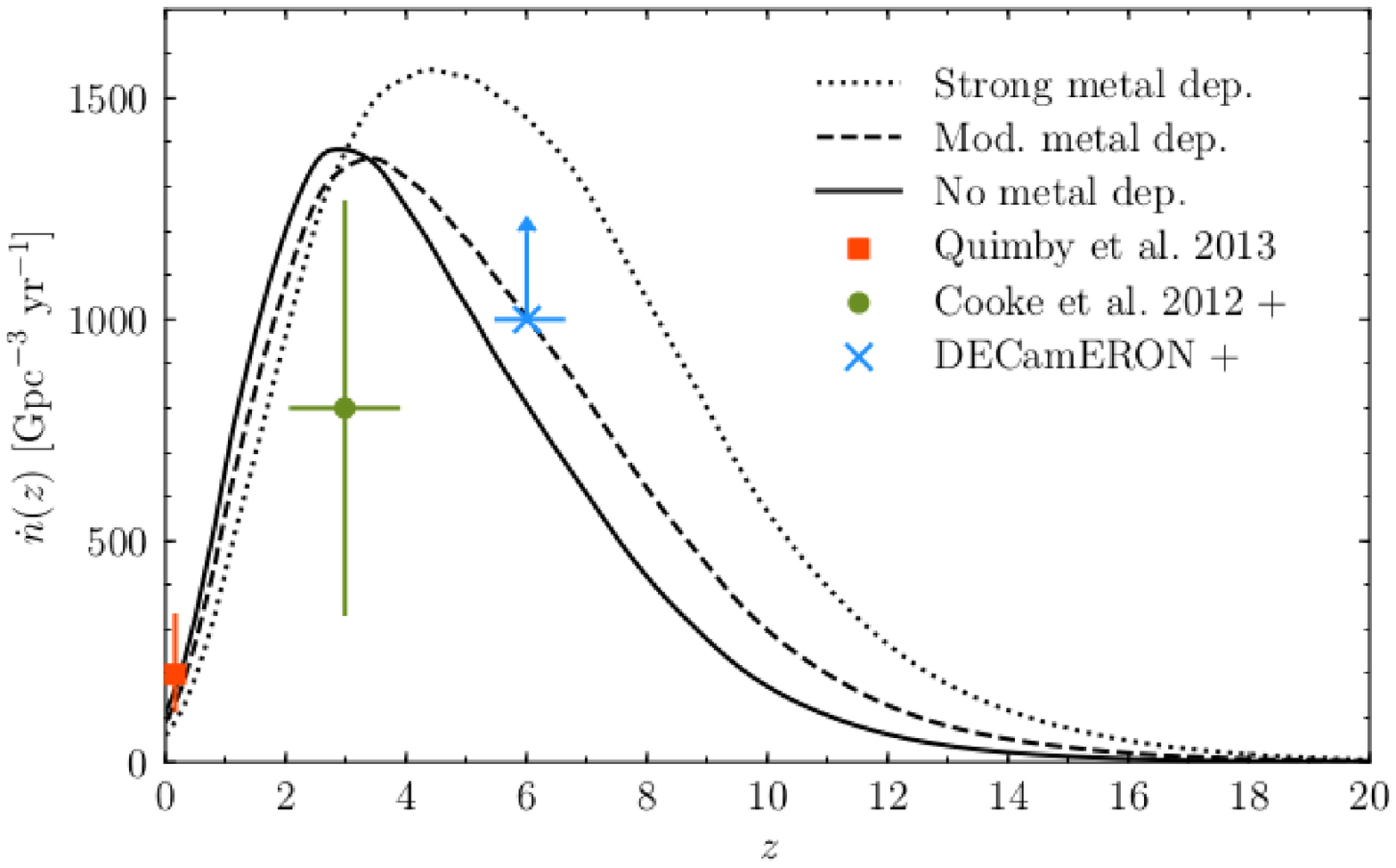}%  SLSN_rate_ERS_v3.eps
\caption{Modelled SLSN rates for three progenitor metallicities. % Top
Above: Volumetric rates normalised to available observations. % (result for 
%$\epsilon_{\rm L} = 0.2$ only shown). %Middle: Observational rates normalised to our survey prediction. Bottom
Below: Integrated observational rates of SLSNe beyond redshift $z$.}

\includegraphics[width=12cm]{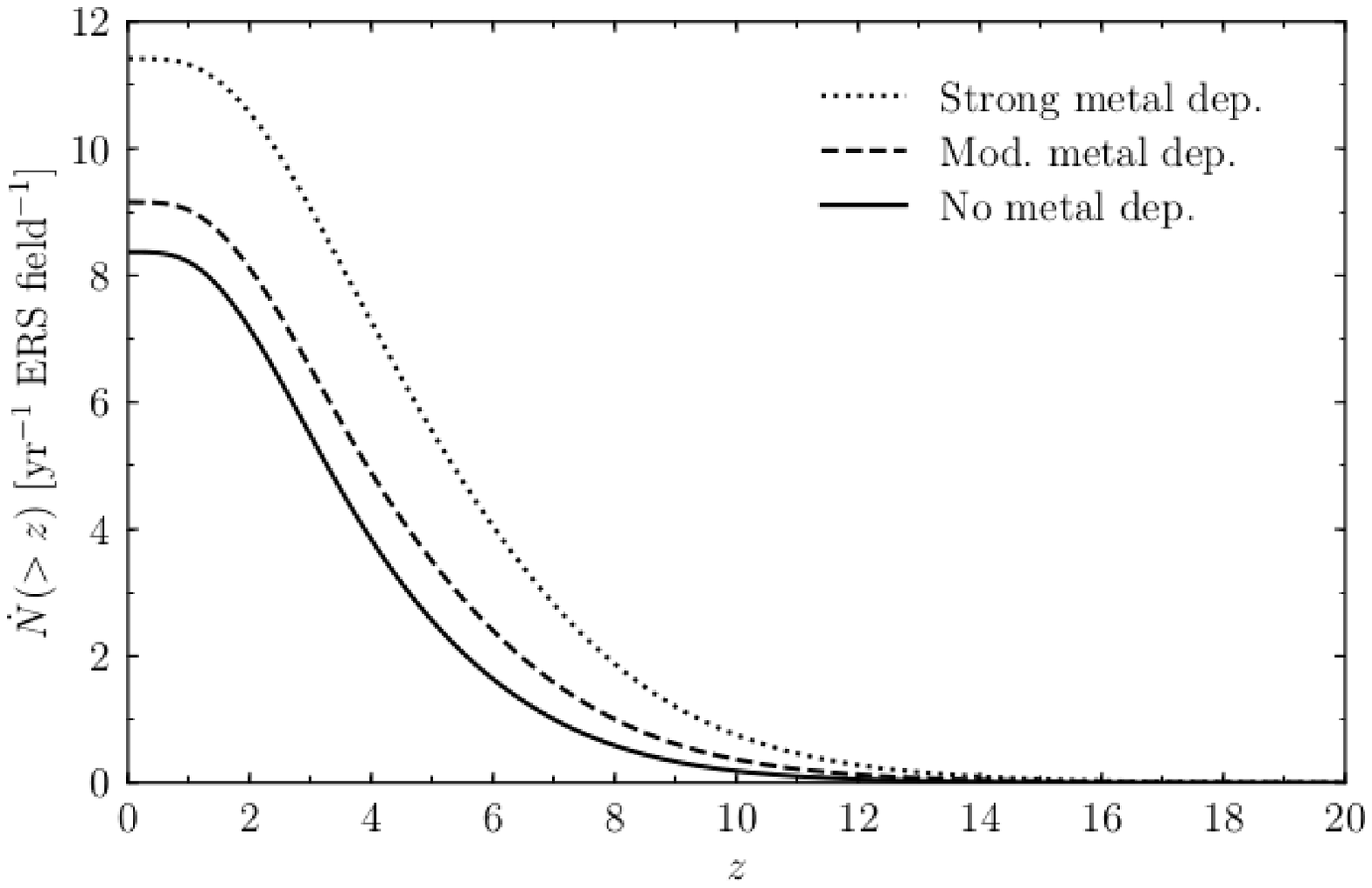}
\label{fig:rates}
\end{figure}

The resulting volumetric rate has been normalised to fit the observed rates of \cite{Quimby2013}, \cite{Cooke2012}, and more recent analysis by Cooke \& Curtin (in preparation). The highest redshift SLSN candidate has been detected by \cite{Mould2017} in deep fields observed with DECam. Their lower limit of 1~deg$^{-2}$ with AB mag $< -22.5$ requires an order of magnitude correction for SLSNe down to $-21.5$ \citep{Quimby2013}. This is included in the upper panel of Figure~\ref{fig:rates} and converted into an observational rate using
\begin{equation}
\dot{N}_{\rm SLSNe} (z) = \frac{\dot{n}_{\rm SLSNe}(z)}{1 + z} \frac{dV}{dz}.
\label{eq:obsrate}
\end{equation}
%This is shown in %the middle panel of 
%Figure~\ref{fig:rates}%, where we have normalised to one SLSN for $6 < z %< 7$ in the ERS field (see Section~2.1.3). 
The lower panel of Figure~\ref{fig:rates} shows the integrated SLSN rate beyond redshift $z$ for three models (strong, moderate and no metallicity dependence). %The predicted observational rates for  three models  ($\epsilon_{\rm L} = 0.2, 0.5$ and a flat efficiency with no metallicity dependence) are similar beyond $z \approx 6$. %We conclude that the expected number of $z \ge 6, 8, 10$ SLSNe is 2.6, 1, and 0.3, respectively.

%In the next section we examine spectral energy distribution (sed) models of SNe. Figure 3 shows similar models for SLSNe.

% \begin{center}
% \end{center}
%\begin{figure}[H]%[ht]
%\begin{center}
%\includegraphics[width=0.7\linewidth]{slsneI_jwst.eps}
%\includegraphics[width=0.7\linewidth]{slsneII_jwst.eps}
%\end{center}
%\caption{SLSN models of type I.}
%\label{fig:SLSNImod}
%\end{figure}
%\subsubsection{ Mid z SLSNe (z$<$6)}
%{\color{red} J Cooke} 

\subsubsection{Normal SNe}
%{\it written by J. Vinko}

\begin{figure}[H]%[ht]
\centering
\includegraphics[width=11.1cm]{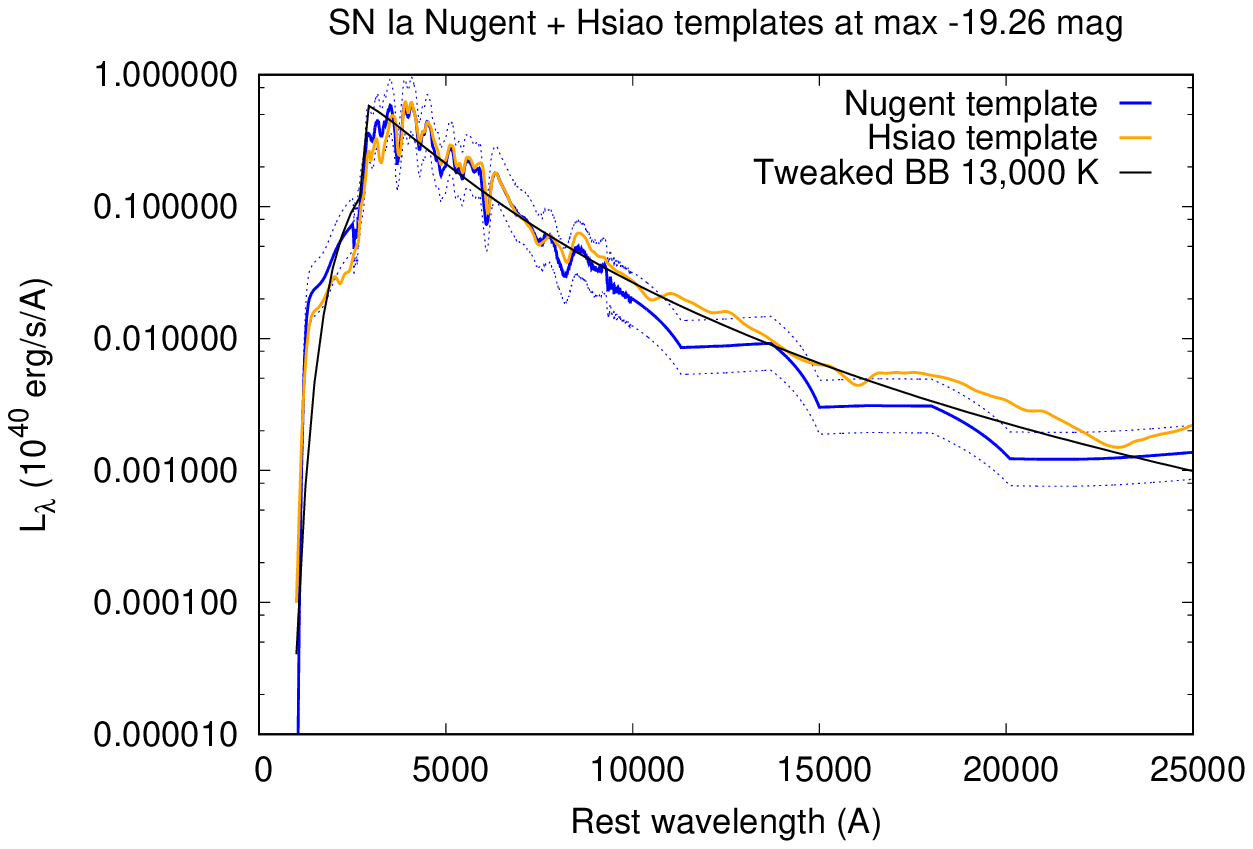}
\includegraphics[width=11.1cm]{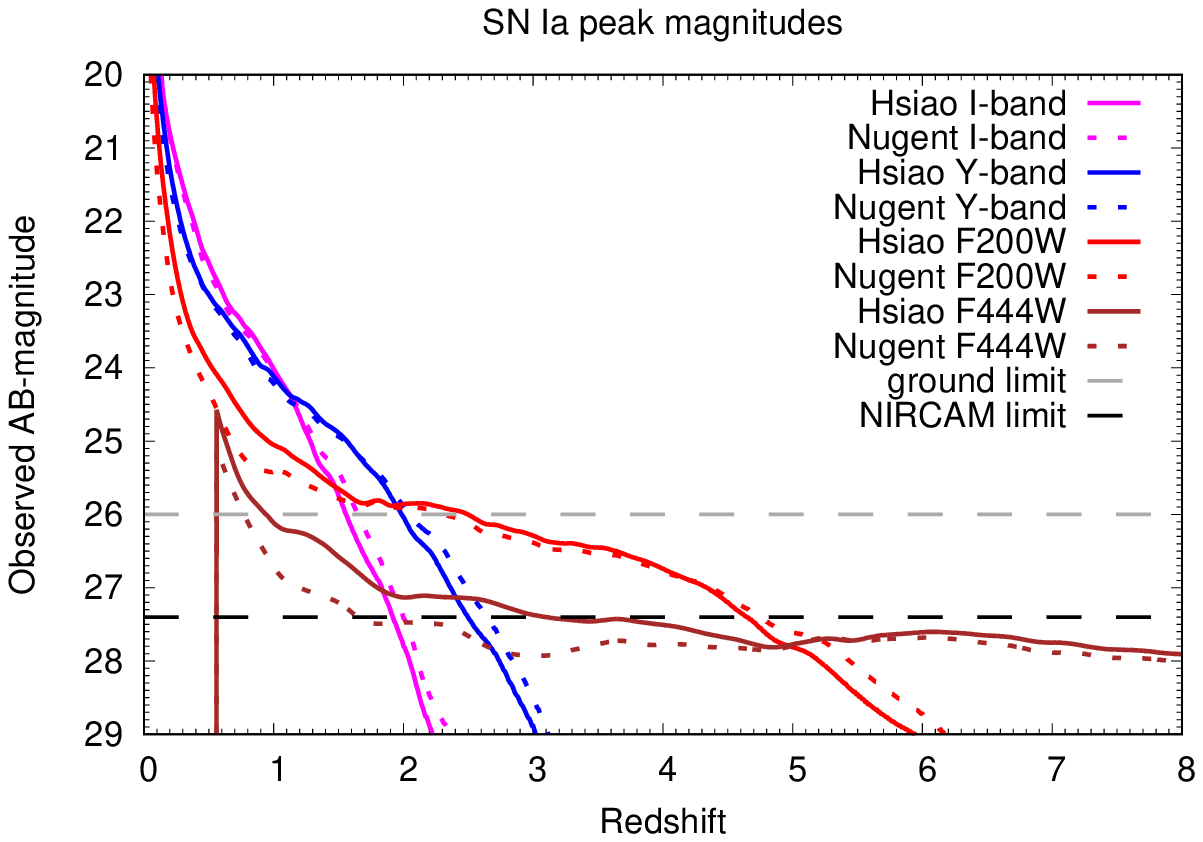}%_highz.eps}
\caption{Left: the Nugent and Hsiao templates for a fiducial Type~Ia SN at maximum with the tweaked black body model (black). 
Right: the predicted AB magnitudes of Type~Ia SNe in various bands as a function of redshift. %(continuous curve: Nugent, dotted curve: Hsiao). 
Dashed horizontal lines indicate the expected sensitivity limit from the ground (grey) and with {\em JWST} (black).}
\label{fig:Ia-max}
\end{figure}

``Traditional'' supernovae, i.e. core-collapse (Type~II and Ib/c) and thermonuclear (Type~Ia) explosions, do not reach sufficient peak brightness to be detected at high ($z > 6$) redshifts. They are abundant, however, in the local ($z < 1$) Universe, and therefore will contaminate the {\em JWST} survey field as the most likely transient sources. Besides the need of filtering out such contaminants from the sample, both CC and Type~Ia SNe could be important in probing the star formation rate (SFR) for $z > 1$. To test the detectability of such kind of SNe with {\em JWST}, we used the various SN spectral templates of \cite{Nugent1997} that extend from 0.1~$\mu$m to 2.5~$\mu$m in wavelength. At first the template closest to maximum light was selected in order to get constraints on the detectability from the maximum distance. Figure~\ref{fig:Ia-max}
shows the Nugent Ia template compared to the more recent Hsiao template \citep{Hsiao2007}  
%scaled to the fiducial maximum magnitude of $-19.26$ Vega-mag. Overplotted with black is a 
tweaked %two-temperature 
black body that was applied to model the approximate behavior of the SED from the UV to the near-IR. This black body fit was applied during the further calculations.

% \begin{center}
% \end{center}

We adopted the cosmological model from the $Planck$ collaboration nicknamed ``Planck13'' in {\tt AstroPy} ($H_0 = 67.77$, $\Omega_m = 0.307$, $\Omega_\Lambda = 0.693$) to calculate the luminosity distances in the redshift range of $0 < z < 10$. The peak brightness (in AB magnitudes) as a function of redshift is plotted in the lower panel of Figure~\ref{fig:Ia-max}.
It is seen that in the $I$- and $Y$-bands SNe~Ia cannot be reached from the ground beyond $z \sim 2$, but this limit can be extended up to $z \sim 5$ with {\em JWST} at 2 and 4.4~$\mu$m.    

% \begin{center}
\begin{figure}[H]%[ht]
\centering
\includegraphics[width=10cm]{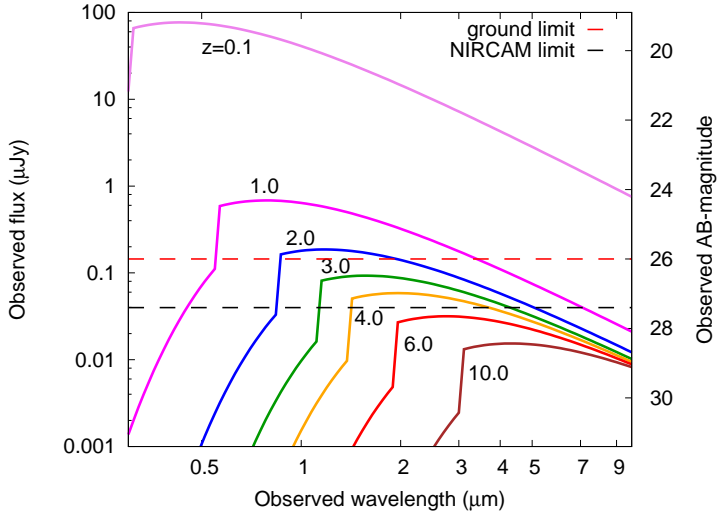}
\caption{The SED of Type~Ia SNe at different redshifts (redshifts are color-coded according to the legend). The figure %left panel
shows the expected flux densities in $\mu$Jy, while the right panel shows the SEDs expressed in AB magnitudes.
The long-dashed horizontal lines illustrate sensitivity limits corresponding to 26 and 27 AB magnitudes.}
\label{fig:Ia-sed}
\end{figure}
% \end{center}

Figure~\ref{fig:Ia-sed} illustrates which wavelength region of the SED of redshifted Type~Ia SNe remains above the detection limits for different redshifts. It is seen that the rest-frame UV/B region of Type~Ia SNe redshifted to $z \sim 4$ --5 are still above the {\em JWST} detection limit.  

These calculations suggest that if a transient (likely a SN) is detected in all 3 bands (1, 2 and 4.4~$\mu$m) at appropriate flux levels then it is likely a Type~Ia SN within the $0 < z < 2$ redshift range. However, if it is detected at 2 and 4.4~$\mu$m but not at 1~$\mu$m then it may be a Type~Ia SN between $2 < z < 5$. Note that detection at 1~$\mu$m is feasible from the ground, so the {\em JWST} survey should be accompanied by a ground-based survey with a sufficiently large telescope (Subaru or Keck). Such observations are not difficult to schedule because time dilation greatly reduces criticality.

Note also that the presence of Type~Ia SNe above $z > 2$ requires the existence of a prompt channel for their progenitors which has negligible delay time after the formation of the carbon-oxygen (C/O) white dwarf. Thus, the {\em JWST} survey can provide very important constraints on the existence of such a prompt channel.

% \begin{center}
\begin{figure}[H]%[ht]
\centering
\includegraphics[scale=0.8]{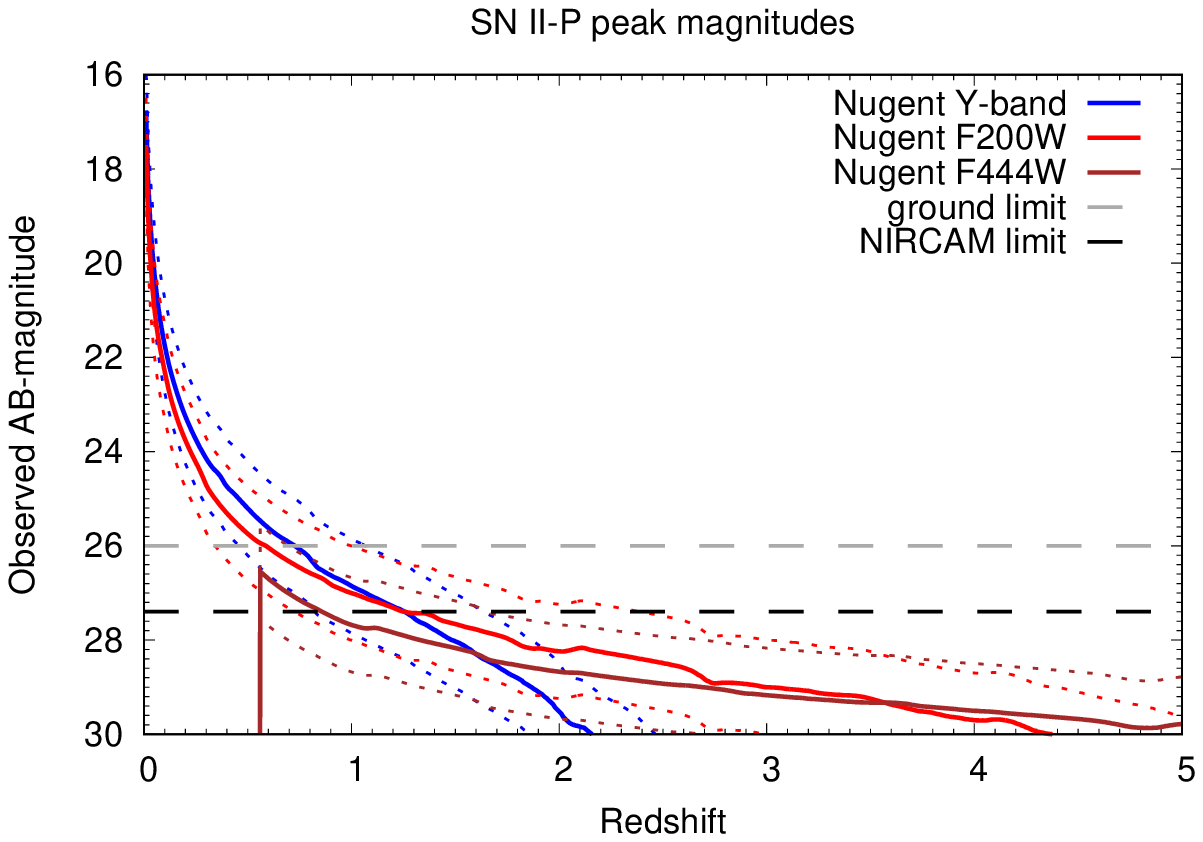}
\includegraphics[scale=0.8]{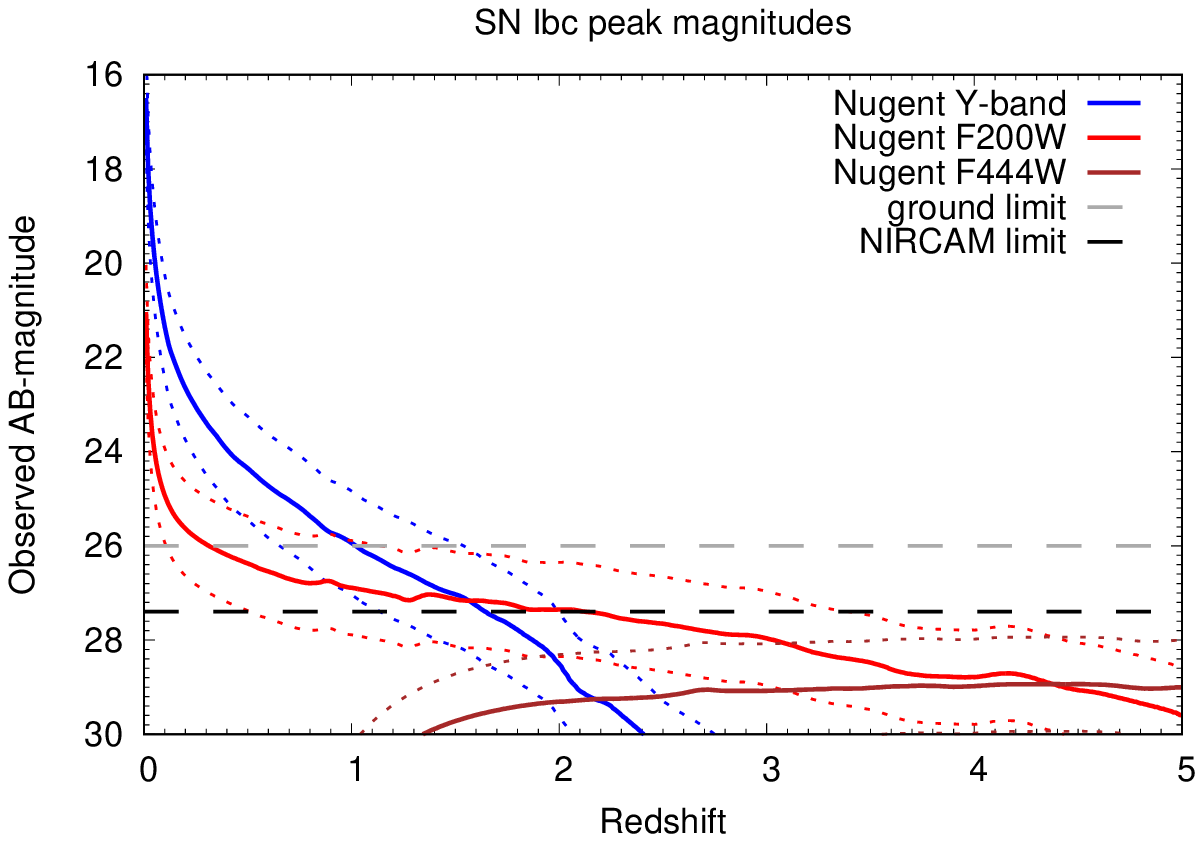}
\caption{Top panel: the range of the expected maximum AB magnitudes of Type~II-P SNe from black body fits to the Nugent templates as a function of redshift. The considered photometric bands are color-coded, and the detection limits from the ground and by {\em JWST} are shown by dashed lines. Lower panel: the same as the top panel but for Type~Ib/c SNe.}
\label{fig:cc-max}
\end{figure}
% \end{center}

Traditional core-collapse SNe (Type II and Type Ib/c) tend to have lower peak brightnesses; thus they are expected to show up only at lower redshift above the detection limit of {\em JWST}. Figure~\ref{fig:cc-max} displays their expected peak brightnesses as a function of redshift, using the Nugent-templates after fitting black bodies to the spectra at maximum.  
%similar to Figure~\ref{fig:Ia-max}. 
Figure~\ref{fig:cc-max} suggests that CC~SNe will be detectable with {\em JWST} only up to $z \sim 2$. 

\begin{table}[H]
\begin{center}
\caption{A possible classification scheme for low-$z$ SNe based on detections in three bands.}
\begin{tabular}{cccl}
Y(1.03~$\mu$m) & F200W & F440W & Classification \\
\hline
+ & + & + & Ia $0 < z < 2.5$ \\
+ & + & + & II-P $0 < z < 1$ \\
+ & + & - & Ibc $z < 1$ \\
- & + & + & Ia $2.5 < z < 5$ \\
- & + & + & II-P $1 < z < 2$ \\
- & + & - & Ibc $1 < z < 3$ \\
\end{tabular}
\label{tab:SNe-classif}
\end{center}
\tablenote{Y band is to be obtained from the ground.}
\end{table}

Table~\ref{tab:SNe-classif} summarizes a possible classification scheme for the low-$z$ SNe based on detections in three bands: the $Y$-band at 1.03~$\mu$m, the {\em JWST} F200W band around 2~$\mu$m and the {\em JWST} F440W band around 4.4~$\mu$m. Plus/minus signs mean detection/nondetection in the given band with fluxes appropriate for a particular SN type.

\subsubsection{SN rates}
The expected number of SNe within the survey field-of-view (FoV) can be calculated: %from Eq.\label{eq:snrate}:
\begin{equation}
N ~=~ \sum_{i} R(z_i) \epsilon_i {T \over {1 + z_i}} dV_i(z_i),
\label{eq:snrate}
\end{equation}
where $z_i$ is the redshift in the $i$-th bin, $T$ is the survey time, $R(z_i)$	is the redshift-dependent SN rate per unit volume, $dV_i$ is the differential comoving volume of the $i$-th redshift bin and $\epsilon_i$ is the survey detection efficiency at redshift $z_i$. The SN rate, $R(z)$, can be expressed as:
\begin{equation}
R(z) ~=~ \int SFH(t(z)-\tau) \cdot DTD(\tau) d\tau,
\label{eq:dtd}
\end{equation}
where $t(z)$ is the cosmic time at redshift $z$, $SFH(t)$ is the cosmic star-formation history at redshift $z$, while $DTD(t)$ is the delay-time distribution for the given SN type. The latter was assumed as a delta function for CC~SNe, meaning that there is no significant time delay between the massive star formation and the following SN explosion. This assumption results in $R(z) = SFH(t(z))$.

\begin{figure}[H]%[ht]
\centering
\includegraphics[width=9cm]{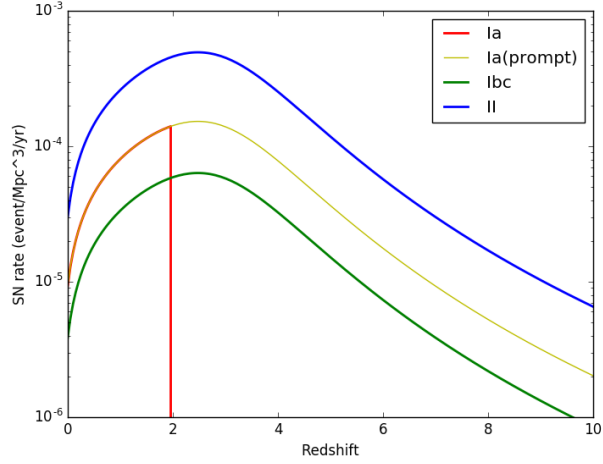}
\caption{The assumed rates for each SN type as a function of redshift.}
\label{fig:snr}
\end{figure}

The rate of Type~Ia SN at high redshifts ($z > 2$) is more uncertain. Current models for Type~Ia progenitors, either the single degenerate or double degenerate scenario \citep{Iben1984}, predict a strongly decreasing Type~Ia rate above $z \sim 2$. There could be, however, a prompt Type~Ia population which might explode very shortly after the formation of the white dwarf (WD). As a first approximation, we assumed that the rate of Type~Ia SN can be smoothly extended to $z > 2$ from lower redshifts, following the cosmic star-formation history function. This approximation gives only a strongly overestimated upper limit for the expected number of Type~Ia SNe at high redshifts. The detected number of such SNe, if any, will be smaller. We applied the empirical formula by \cite{Hopkins2006} to estimate the $SFH(z)$ function at different redshifts. Values for the absolute SN rates of the particular types were collected from \cite{Bazin2009}, while the fractions of Types~II and Ibc within the CC~SNe were adopted from \cite{Shivvers2017}, being 0.7/0.3. Figure~\ref{fig:snr} 
shows the adopted SN rates for each type.

% \begin{center}
% \end{center}

Finally, the number of SNe within the survey FoV was calculated from Eq.~\ref{eq:snrate} for each type. We adopted $T = 100$~days for the total survey time %during the ERS 
and the following detection limits in the three bands: $Y$ (from ground): 26 AB mag; $F200W$ ({\em JWST}): 27 AB mag; $F440W$ ({\em JWST}): 27 AB mag. Assuming 100\% survey efficiency ($\epsilon_i = 1$) as a first approximation, the results are collected in Table~\ref{tab:snnum}. 

\begin{table}[H]
\begin{center}
\caption{The expected number of SNe of various types in the survey FoV during the survey time}
\begin{tabular}{cccccccc}
SN type & $M_{peak}$ (V) & N (Y) & N (F200W) & N (F440W) \\
\hline
Ia (w/o prompt) & $-$19.3 & 19 & 19 & 13 \\
Ia (w prompt) & $-$19.3 & 23 & 45 & 13 \\
Ibc & $-$17.6 & 5 & 6 & 0 \\
II & $-$16.8 & 4 & 16 & 4 \\
\end{tabular}
\label{tab:snnum}
\end{center}
\tablenote{100\% detection efficiency assumed.}
\end{table}

In total, we can expect $\sim 45$ Type~Ia, $\sim 16$ Type~II and $\sim 6$ Type~Ibc SNe to be detectable in the survey field. The number for Type~Ia SNe is an upper limit, as it contains the overestimated fraction of the prompt Type~Ia SNe. Without the prompt population, the expected number of Type~Ia SNe reduces to 19. 

\subsubsection{Finding the First Type~Ia SNe}%Thermonuclear Supernovae}
The %ERS 
survey we propose will be able to discover thermonuclear supernovae to a redshift approaching z $\sim$ 4. Although the sparse light curve sampling from our survey may not allow us to accurately extract the light curve shapes and derive precise cosmological distances to these SNe, they are nonetheless important in constraining the progenitor systems of Type~Ia SNe and test the evolution of their intrinsic magnitudes. Type~Ia SNe at such redshifts will experience time dilation by a factor of 4--5. Once discovered, they are expected to be visible by {\em JWST} for about 6 months, and bright enough for photometric and spectroscopic followup. They can be studied by separate programs aiming to directly measure cosmic deceleration at those redshifts, the ISM towards these SNe, and stellar evolution at  very high redshifts. Because we will also discover $\sim 50$ Type~Ia SNe at redshifts around $z \sim 2$, our %ERS 
program will serve as target feeder to programs aiming to measure precision cosmological parameters.

\subsection{Large Scale Structure in the Epoch of Reionization}
%{\it written by Jeremy Mould}

\begin{figure}[H]%[ht]
\centering
\includegraphics[angle=-90,width=0.8\textwidth]{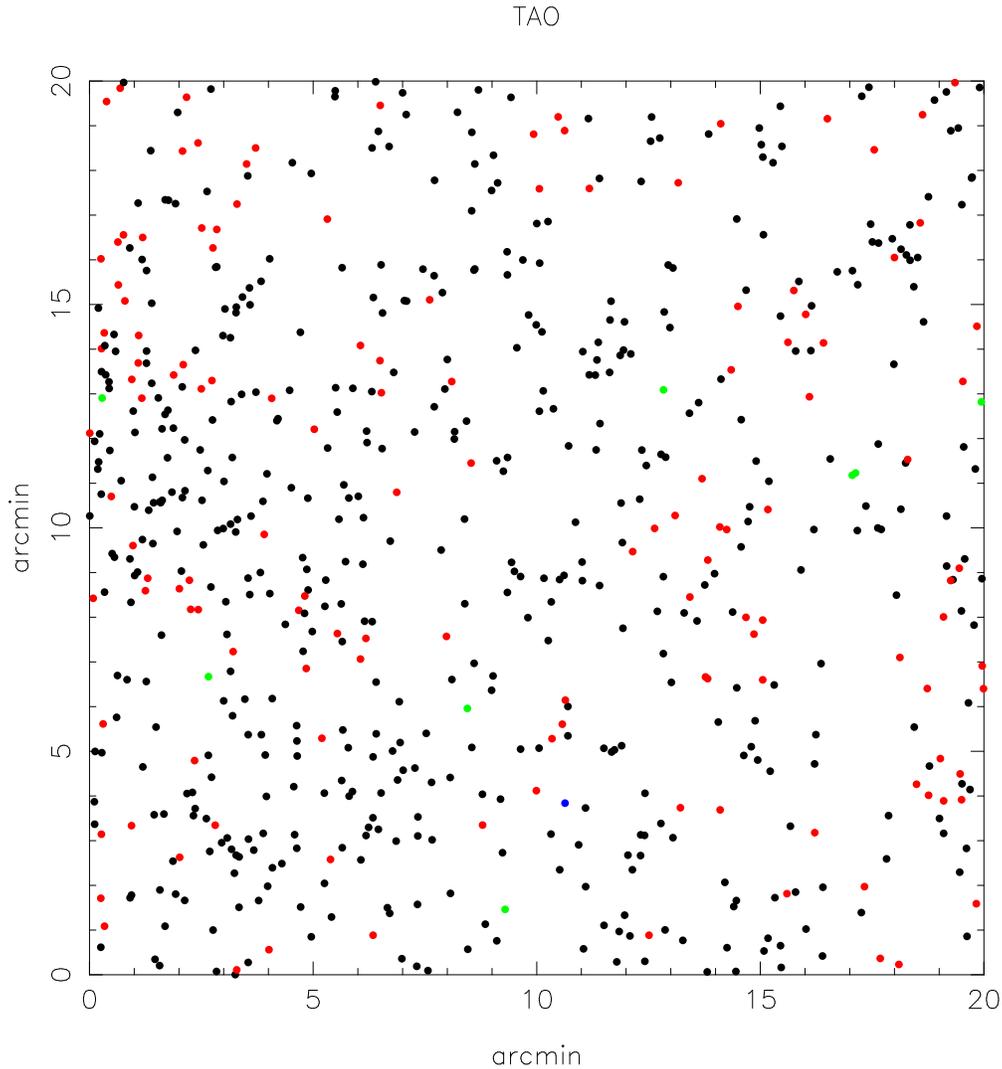}
\vspace{-2 truemm}
\caption{Brightest galaxies in a light cone made with TAO in a 400~square arcminute field. Black points have $z = 6$; red points $z = 7$ to blue $z = 9$; %green z = 8.
One comoving Mpc is $\approx 3$~arcminute at $z = 6.56$. The higher density of galaxies in the centre left of the field will yield the largest intensity of ionizing radiation and thus blow a bubble in the neutral hydrogen gas, as seen in the DRAGONS simulations.}
\label{fig:TAO}
\end{figure}

\begin{figure}[H]%[ht]
\centering
\includegraphics[angle=-0,width=0.7\textwidth]{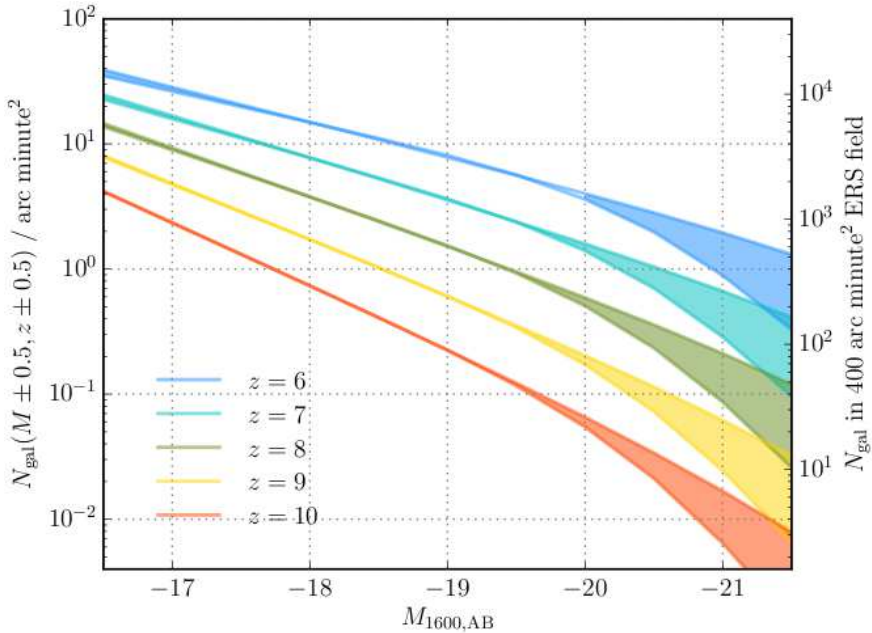}
\vspace{-2 truemm}
\caption{The number of galaxies are bound at each redshift by the intrinsic UV luminosity function (upper curve of the shaded envelope) and the dust-corrected UV luminosity function (lower curve of the envelope). Dust is exaggerated at $z = 9$--10. Integrated in $dz = 1$ and $dM = 1$ bins centered over $z$ and $M_{\rm UV}$ ($M_{\rm 1600,AB}$).}
\label{fig:LF}
\end{figure}

An important question in the epoch of reionization (EoR) is, can a change in the power spectrum of the spatial distribution of galaxies be detected between $z = 6$--10? %redshift 10 and redshift 6?
We consider whether this is possible in a 0.1~square degree FLARE field.
%Dropout galaxies are redder than brown dwarfs, as can be seen from the next two figures contributed by Travis Barman. 
A typical distribution of galaxies in the light cone between $6 < z < 10$ can be created using the Theoretical Astrophysical Observatory\footnote{\url{https://tao.asvo.org.au}} \citep[TAO;][]{Bernyk2016} and the Millennium simulation (\citealt{Harker2006}, plus SAGE semianalytics; \citealt{Croton2016}). The brightest galaxies are shown in Figure~\ref{fig:TAO}. The luminosity function is shown in Figure~\ref{fig:LF}.

The number of galaxies is made to equal the predictions of the DRAGONS simulation \cite{DRAGONS3}.

\subsection{Growing supermassive black holes before Reionization}
%{\it written by Fabio Pacucci and Bin Yue}\\

The first SMBH seeds\footnote{Meaning BH beyond stellar mass} formed when the Universe was younger than $\sim 500$~Myr and played an important role in the growth of early ($z \sim 7$) SMBHs \citep[e.g.][]{Pacucci2015,Natarajan2017}. Much progress has been made in recent years in understanding their formation, growth and observational signatures, but many questions remain unanswered and we are yet to detect these sources. \cite{Natarajan2017} predicted the observational properties and {\em JWST} detectability of black hole seeds, formed by the direct collapse of high-redshift halos (e.g. \citealt{bl03}, or as remnants of Pop~III stars, e.g. \citealt{vr05}). 

When primordial, atomic-cooling halos ($T_{\rm vir}>10^4 \, \mathrm{K}$) are exposed to a high-intensity Lyman-Werner flux, $J_{\nu} > J_{\nu}^{\bullet}$ \citep{Loeb_Rasio_1994,Lodato_Natarajan_2006, Shang_2010} the destruction of $\rm H_2$ molecules allows a rapid, isothermal collapse. The precise value of $J_{\nu}^{\bullet}$ depends on several factors, but there is a general consensus that it should fall in the range $30 < J_{21}^{\bullet} < 1000$, depending on the spectrum of the sources \citep{Sugimura_2014}. Several theoretical works \citep{Bromm2003, Begelman_2006, Volonteri_2008, Shang_2010, Johnson_2012} have shown that the result of this collapse is the formation of a direct collapse black hole (DCBH) of mass $M_\bullet \approx 10^{4-6} \, \mathrm{M_{\odot}}$ \citep{Woods2017,Umeda2016}).

Guided by theoretical estimates in \cite{Pacucci2015} and \cite{Natarajan2017}, these sources are predicted to be particularly bright in the infrared. In the near infrared, these sources should be significantly brighter than 28th magnitude (26 on average, depending on the initial BH  mass and on the physical properties of the host halo), while Pop~III seeds are characterized by infrared magnitudes above 29 on average and they could be unobservable with the {\em JWST}. %\citep{Natarajan2017}. 
\cite{Pacucci2016} \citep[see also][]{pal15,Pacucci2017} claimed the possibility of two $z\gapp 6$ DCBH candidates in a survey such as CANDELS/GOODS-S \citep{Illingworth2016} with a significant X-ray emission. %These sources are predicted to be bright at wavelengths > 1~$\mu$m, while at < 0.9~$\mu$m they could be unobservable due to IGM absorption, depending on the specific redshift of the source. 
DCBHs are thus high-value targets for the {\em JWST} in the infrared. In what follows we present a general overview on the properties and detectability of DCBHs.

\subsubsection{The brightness, number density and detectability of DCBHs}

Assuming that a DCBH is accreting at the Eddington rate and that it is Compton-thick \citep{Yue2013} ($N_{\rm H}\gapp 10^{24} $~cm$^{-2}$, where $N_{\rm H}$ is the column number density of the host galaxy), then its 4.5~$\mu$m apparent magnitude is a simple function of black hole mass. % is shown in %Figure~\ref{fig:magDCBH}.
The computation is performed for an object located at $z \sim 13$, i.e. well inside the cosmological period during which the formation of DCBHs is more likely \citep{Yue2013}. A DCBH seed is predicted to have a mass in the range $\sim 10^4$--$10^6$~M$_\odot$ \citep{Ferrara2014}. When it grows to $\gapp 2\times10^6$~M$_\odot$, it becomes detectable in the survey that we are proposing (i.e. brighter than 26.5 mag at 4.5~$\mu$m). A DCBH could, however, be characterized by super-Eddington accretion rates \citep{Pacucci2015}. In this case the relation between mass and brightness is less straightforward to predict, and a DCBH could be detectable even for masses $\lapp 2\times10^6$~M$_\odot$.

%\begin{figure}[H]%[h]
%\centering
%\includegraphics{[width=0.8\textwidth]{magDCBH.eps}
%\includegraphics[width=0.6\textwidth]{magDCBH.eps}%{mag_DCBH.pdf}
%\caption{\label{fig:magDCBH}The 4.5~$\mu$m apparent magnitude of a %Compton-thick DCBH vs. 
%the black hole mass, for a DCBH at $z \sim 13$.}
%\end{figure}

\begin{figure}[H]%[ht]
\centering
\includegraphics[width=0.6\textwidth]{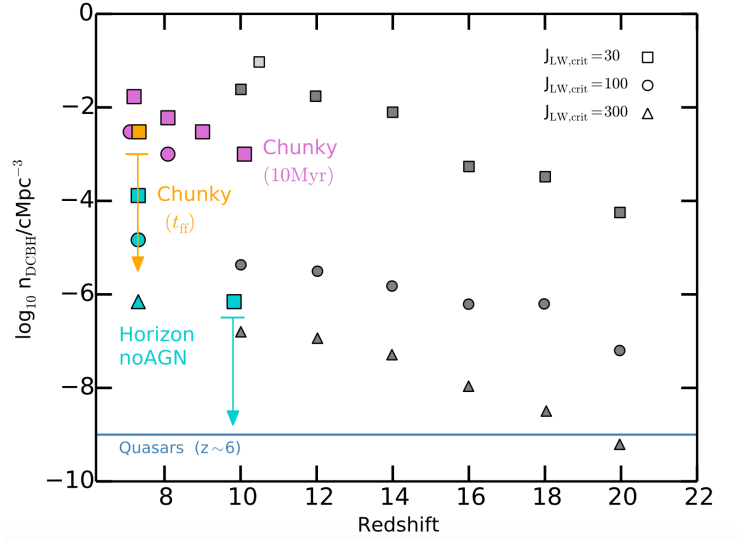}%{numDCBH.pdf}
\caption{The number density of DCBHs predicted by several works . The figure is adapted from \cite{Habouzit2016}. While the detailed description of each model is found in the caption of the original paper (and tabulated below), here we only choose to show the great variety of predictions.}
\label{fig:numDCBH}
\end{figure}

In the literature, the predicted number density of DCBHs widely varies, from $\sim 10^{-10}$--$10^{-1}$~Mpc$^{-3}$ at $z \sim 10$ (comoving units). The large span in the predictions is mainly due to the uncertainties on the critical external field strength that can fully suppress the $\rm H_2$ formation, the clustering of the potential DCBH formation sites, and the feedback effects. A summary of the DCBH number densities predicted by different papers is shown in Figure~\ref{fig:numDCBH} \citep[taken from][]{Habouzit2016}. While the detailed description of each model is found in the caption of the original paper, here we just want to show the great variety of predictions. The number density could be even as high as $\sim 0.1$~Mpc$^{-3}$ at $z \sim 13$, due to our condition of producing the observed cosmic near-infrared background %(NIRB) 
fluctuations levels \citep{Yue2013} from these sources.

\begin{table}[H]
\centering{
\caption{Symbols and models.}
  \begin{tabular}{llllll}
      Symbols &$n_{\rm DCBH}$~Model & $M_\bullet^*$& $N_{\rm H}$ \\ \hline\hline
       stars &Y13 &  \--- & $1.2\times10^{25}$~cm$^{-2}$ \\      
      filled circles &A12, $J_{\rm LW}^{\rm c}=30$& $10^6$~M$_\odot$ & $10^{25}$~cm$^{-2}$ \\  
       open circles &D14, $J_{\rm LW}^{\rm c}=30$& $10^6$~M$_\odot$ & $10^{25}$~cm$^{-2}$ \\  
       
        filled squares &A12, $J_{\rm LW}^{\rm c}=30$& $5\times10^5$~M$_\odot$ & $10^{25}$~cm$^{-2}$ \\  
       open squares &D14, $J_{\rm LW}^{\rm c}=30$& $5\times10^5$~M$_\odot$ & $10^{25}$~cm$^{-2}$ \\  

       filled triangles &A12, $J_{\rm LW}^{\rm c}=30$& $10^6$~M$_\odot$ & $10^{20}$~cm$^{-2}$ \\  
       open triangles &D14, $J_{\rm LW}^{\rm c}=30$& $10^6$~M$_\odot$ & $10^{20}$~cm$^{-2}$ \\  
                  
       \hline
  \end{tabular}
  \label{tb_surface_density}
  }
\end{table}

\begin{figure}[H]%[ht]
\centering
\includegraphics[width=0.7\textwidth]{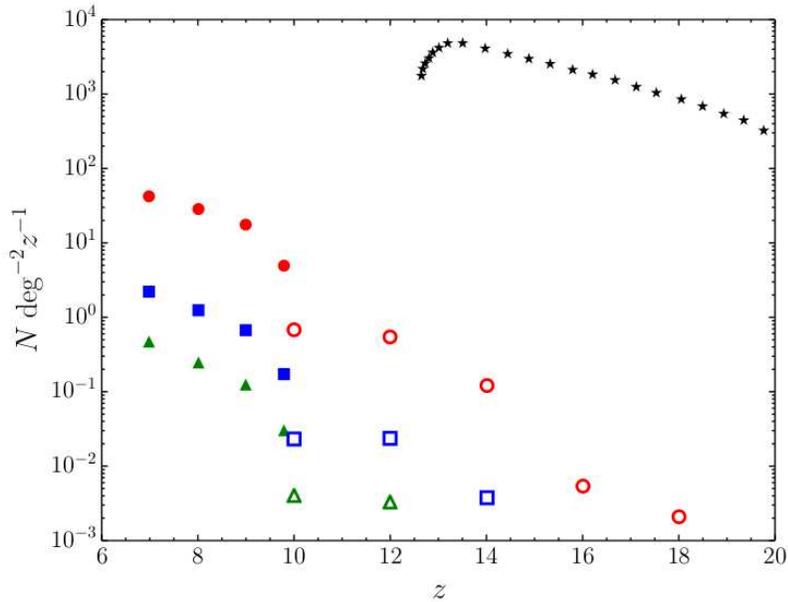}
\caption{The surface number density of DCBHs brighter than 26.5~mag at 4.5~$\mu$m in various models. We show the surface number densities predicted in \cite{agarw12}, \cite{Yue2013} and \cite{Dijkstra2014} for various DCBH and mass function parameters.}
\label{fig:surfaceDCBH} 
\end{figure}

Regarding the detectability, we convert the number density to a surface number density of DCBHs brighter than some threshold magnitude by specifying a mass function. Notwithstanding large uncertainties in the DCBH mass function, we can assume a Schechter formula or a bimodal Gaussian distribution \citep{Ferrara2014}. In Figure~\ref{fig:surfaceDCBH} we show the surface number density of DCBHs brighter than 26.5 mag at 4.5~$\mu$m, for the DCBH number density predicted in \cite{agarw12} (A12), \cite{Yue2013} (Y13) and \cite{Dijkstra2014} (D14) for various DCBH and mass function parameters (see Table~\ref{tb_surface_density}). Basically, Compton-thick DCBHs are more detectable because more energy is re-processed to the rest-frame UV/optical band, then redshifted to the near-infrared band. 

%Multiplying the surface number density of a steady accreting DCBH (Fig. %14) by this rate, we obtain the detectability of tidal disruption events %(TDEs): $\sim 10^{-3}$ deg$^{-2}z^{-1}{\rm yr}^{-1}$ at maximum at z %$\sim 13$ 
%in Y13, and $\sim 10^{-5}$~deg$^{-2}z^{-1}{\rm yr}^{-1}$ at maximum at z %$\sim 7$ in A12. 
%However, if the TDE happens through the whole lifetime of a DCBH, then %we expect a rate as high as 0.05 deg$^{-2}z^{-1}{\rm yr}^{-1}$ in Y13.

\subsubsection{The infrared colors of DCBHs}

According to \cite{Pacucci2016}, a DCBH has infrared colors significantly different from the typical QSO or star-forming galaxy. For the photometric filters of our interest, Y (1~$\mu$m), 2.2~$\mu$m and 4.4~$\mu$m, we predict photometric colors $Y-2.2$ and $2.2-4.4$ larger than 2. %, in accordance with the %significant redness of these sources.  
DCBH candidates can be preselected even in purely photometric surveys.

%\subsubsection{Observational features of DCBH candidates}
Spectroscopic signatures of DCBHs are: (i) strong He\,II 1640~\AA~emission line, (ii) strong Ly$\alpha$ emission (but if the DCBH is extremely Compton-thick, the Ly$\alpha$ emission would be trapped and converted into continuum emission), (iii) absence of metal lines. Candidates could also be selected according to their photometric colors (Figure~11) \citep{Pacucci2016}. 
%X-ray observations are nonetheless necessary to confirm the candidate. 
%However if a DCBH is extremely Compton-thick, in principle X-rays could %only be detected in the hard band. 
%Even in the most pessimistic case, the {\em JWST} at least could provide a %catalog of DCBH candidates.
%The lack of metal lines is not strictly necessary for identifying a DCBH %candidate, 
%since the leftover gas from the formation of the DCBH can form stars and %generate metal %\citep{Agarwal2013, Natarajan2017} 
%(Agarwal et al 2013, Natarajan et al 2017).

\begin{figure}[H]%[ht]
\centering
\includegraphics[width=.7\textwidth]{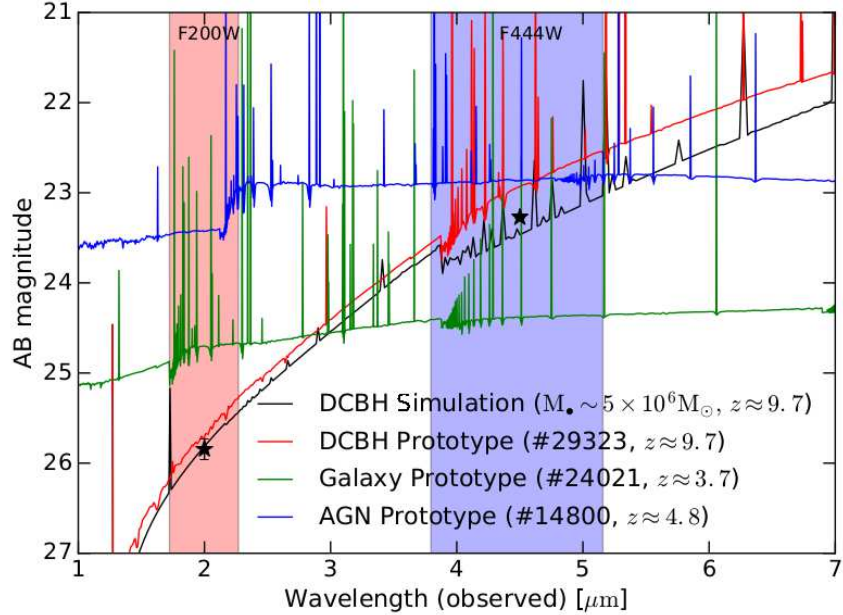}
\caption{Comparison between the stellar SEDs of three GOODS-S objects with the computed SED of a $5 \times 10^6$~M$_\odot$ black hole, born out of a DCBH with initial mass around $10^5$~M$_\odot$. NIRCAM photometric bands are shaded. Stars (numerical simulations) show the magnitude %in the three filters, with 
error bars. Objects 29323 and 14800 have X-ray counterparts (i.e. they are likely associated with a black hole), while 24021 does not (i.e. it is likely a normal galaxy). Moreover, object 29323 is characterized by very negative colors (i.e. its infrared SED is very steep, as we predict for DCBHs), while objects 14800 and 24021 are not. The steepness of the SED and the infrared magnitudes for the object 29323 are well fitted by the spectrum predicted for a $\sim 5 \times 10^6$~M$_\odot$ black hole. In the computed SED for a DCBH, the He~II line (0.164~$\mu$m rest-frame) is visible and it is marginally inside the H~band at $z \sim 9.7$.}
\label{fig:seds}
\end{figure}

\begin{figure}[H]%[ht]
\centering
\includegraphics[width=.5\textwidth]{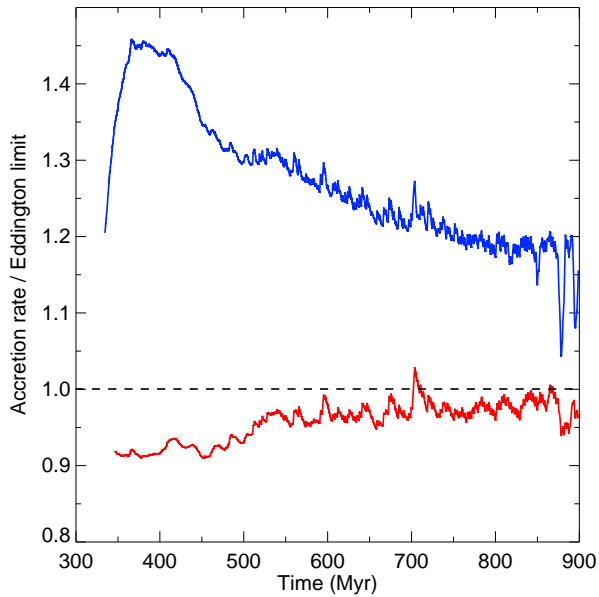}
\caption{DCBH accretion rates as a fraction of the Eddington limit. Blue: no X-ray feedback. Red: with X-ray feedback from the BH.}
\label{fig:accr}
\end{figure}

Although TDEs around DCBHs produce spectacular transients that are well within the sensitivity of our proposed survey, fluctuations in normal accretion onto the DCBH also drive variations in near-infrared (NIR) luminosities that could be detected as well. These variations are due to disruptions in flows onto the DCBH over a large range of spatial scales and times. New radiation hydrodynamical simulations of DCBH growth in cosmological environments show that fluctuations in cold flows into the host galaxy on kpc scales can lead to large variations in BH luminosity on timescales of Myr, as shown in Figure~\ref{fig:accr} \citep{smidt17}. But brightening and dimming could occur on times as short as the light-crossing time of the BH. Numerical simulations that achieve sub-AU resolution show that catastrophic baryon collapse in atomically-cooled halos leads to the formation of bursty accretion disks around supermassive primordial protostars, the precursors to DCBHs \citep{bec15}. Clumpy accretion due to turbulence in the disk can result in changes in luminosity of about an order of magnitude on timescales of days or weeks in the rest frame of the nascent DCBH. Such variations should be easily detectable by our proposed survey.

%In addition to the DCBH in steady accretion state, the tidal disruption event (TDE) 
%is also worth of consideration in the survey strategy. 
%Returning to the subject of TDEs, a
According to \cite{Kashiyama2016}, the TDE rate is $\sim 10$ per DCBH within $\sim 1$~Myr at the DCBH early growth stage. If the TDE only happens at the early stage, and assuming that the typical lifetime of a DCBH is $\sim 50$~Myr, then we expect the TDE rate to be $\sim 2 \times 10^{-7}$~yr$^{-1}$ per DCBH. Multiplying the surface number density of a steady accreting DCBH (Figure~\ref{fig:surfaceDCBH}) by this rate, we obtain the detectability of TDEs: $\sim 10^{-2}$~deg$^{-2}z^{-1}$yr$^{-1}$ at maximum at $z \sim 13$ in Y13, and $\sim 2\times10^{-3}$~deg$^{-2}z^{-1}$yr$^{-1}$ at maximum at $z \sim 7$ in A12. However, if the TDE happens through the whole lifetime of a DCBH, then we expect a rate as high as $\sim 0.5$~deg$^{-2}z^{-1}$yr$^{-1}$ in Y13. This value, while obviously lower when compared to the rate of TDEs for all SMBHs, %it 
is %still worth of 
a serious consideration for the present survey. A signature of 
%these is %
TDEs is %, manifesting themselves as 
order of magnitude flares in rise times of $(1+z)$ times 30 days.

%\subsubsection{Tidal disruption events}

%Regarding the TDE events, I didn't see the prediction; however  https://arxiv.org/pdf/1602.04293.pdf uses a reference number 10 per DCBH. If we assume the typical accreting time of a DCBH is 50 Myr, then for each DCBH the TDE rate is 2 x 10$^{-7}$ yr$^{-1}$. Multiply this number with the detection rate of DCBH at steady accretion phase shown in the figure, and we can get the TDE rate. The TDE rates thus appears to be too small to be detectable {\it says Bin Yue}

\subsection{Serendipitous high-redshift transients}

\subsubsection{Kilonovae}
%{\it written by Ed Baron}

\begin{figure}[H]%[h]
\centering
\includegraphics[scale=0.5]{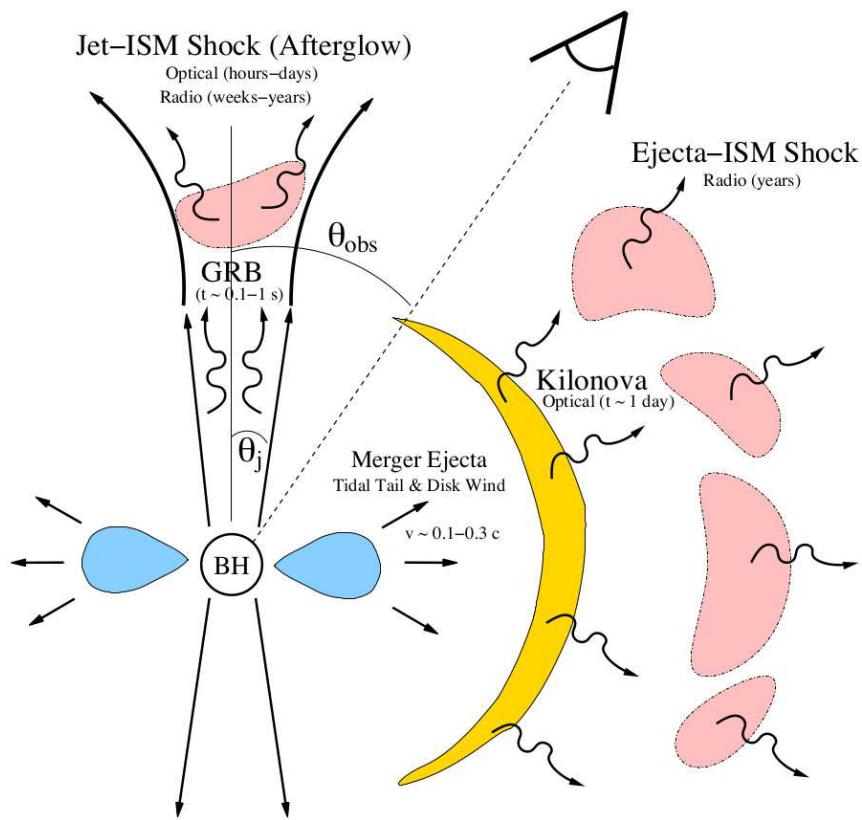}
\caption{The basic picture of the various electromagnetic signatures produced in a merging neutron star event. The kilonova emission comes from the semi-symmetric ejecta \citep[from][]{Metzger2012}.}
\label{fig:kilonova_cartoon}
\end{figure}

The tidal disruption of a neutron star in a binary companion with either another neutron star or a black hole has long been of astrophysical interest, since it has long been understood that the decompression of neutron star material could be a site for r-process nucleosynthesis \citep{Lattimer1974, Lattimer1977, Symbalisty1982}. Interest was %has been
further heightened when it was realized that the same systems could provide an electromagnetic counterpart for gravitational wave signals \citep{Li1998, Kulkarni2005, Metzger2010}. These systems have been dubbed kilonovae or macronovae, but we will use the former name. For an excellent review of the entire field, see \cite{Metzger2017}. The expected electromagnetic signatures are illustrated in Figure~\ref{fig:kilonova_cartoon}. 

\begin{figure}[H]%[ht]
\centering
\includegraphics[width=12cm]{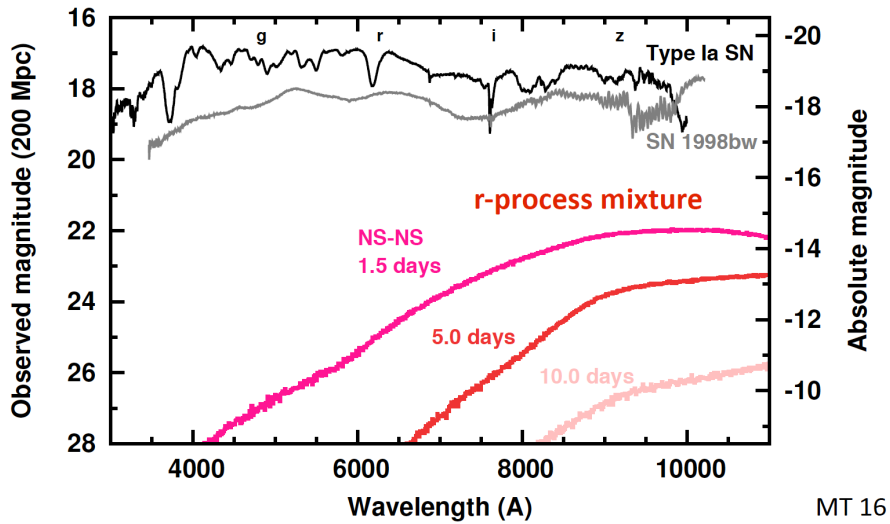}
\caption{The spectrum of the relativistic ejecta of a NS--NS merger compared with that of supernovae \citep[from][]{Tanaka2016}.}
\label{fig:tanaka_spec}
\end{figure}

Population synthesis models predict gravity wave detection rates of NS--NS/BH--NS mergers of $\sim 0.2$--300 per year, for the full design sensitivities of Advanced LIGO/Virgo \citep{Abadie2010,Dominik2015}. Empirical estimates predict $\sim 8$ NS--NS mergers per year in the Galaxy \citep{Kalogera2004a, Kalogera2004b, Kim2015}. Numerical models of the event rate estimate $\sim 1000$~Gpc$^{-3}$yr$^{-1}$ \citep{Tanaka2016} with a spectrum shown in Figure~\ref{fig:tanaka_spec}.

Models of the disk-wind structure have been performed by \cite{Kasen2015} and the dynamics and nucleosynthesis of BH--NS mergers have been studied by \cite{Fernandez2017} and \cite{Metzger2017} and references therein.

We have begun preliminary modeling with the generalized stellar atmospheres code \phx (P.~Vallely \& E.~Baron, in preparation). We include some r-process elements in full NLTE, and the time is right to begin NLTE modeling as the atomic data to construct the model atoms for all r-process elements is now available \citep{Fontes2017}.

Several kilonovae should be visible in our proposed program and understanding these events is important for understanding the site of the r-process as well as crucial to providing eletromagnetic counterparts to gravitational wave events.

\subsubsection{Formation of globular clusters}
%{\it written by Jeremy Mould}\\
%SLSNe and the formation of globular clusters  

\cite{Renzini2017} suggests that a proto globular cluster in the EoR reaches AB = 28~mag for a Myr. At low metallicity these objects are candidates to produce SLSNe with $25 < {\rm AB} < 28$. %Renzini almost ignores extinction, but at the current epoch we detect Super Star Clusters in the IR . 
If 5\% of the cosmic SFR from $z =$~9 to 6 were in globular clusters (GCs), we would have $2 \times 10^5$~M$_\odot$Mpc$^{-3}$. For $M/L = 10^{-2}$ that is $2 \times 10^7$~L$_\odot$Mpc$^{-3}$. In 1~square degree, $dn/dt \sim 1$ GC SLSN per year.

Globular clusters are laboratories for SNe. Since IMFs $\propto m^{-2}$, some half of the mass terminates in SNe. The kinetic energy of a $10^6$~M$_\odot$ GC is $10^6 \times 10^{12} \times 10^{33} = 1$~foe. Depositing 1~foe in it from a SLSN will come close to unbinding the cluster. According to chemical evolution theory removing the gas from the cluster is {\em required} to retain the observed low metallicity, and a SLSN may well be the agent\footnote{See also \cite{Recchi2017},\cite{Boylan-Kolchin2017}
.}.

%\section{A 0.1 Square Degree Field with {\em JWST}}
\section{The FLARE {\em JWST} Transient Field}
\label{section:FLARE}

Now we need to detail the methodology for an efficient {\em First Transients} survey. We propose to survey the North Ecliptic Pole (NEP) %and the South Ecliptic Pole (SEP) 
in two colors down to 27.4 mag (AB). This is very much deeper than the {\em JWST} Time-Domain Community Field \citep{Jansen2017}. %The total survey area consists of mosaics of 3$\times$7 and 3$\times$8 {\em JWST} pointings at the NEP and SEP respectively. 
The NIRCAM with filters F200W and F444W is most appropriate to the project. We will also employ NIRISS for deeper coverage in the F444W band for a small fraction of the field in the parallel mode. For an Early Release Science time-domain survey, two visits of the NEP field would be made at two epochs separated by 91.3~days. For these purposes the primary survey instrument is NIRCAM, with NIRISS in parallel mode. The survey employs two filters, F200W and F444W with NIRCAM and F444 only with NIRISS. Medium background level is appropriate. The detector is to be set up to read out the full array, and the readout pattern is SHALLOW4; the observations employ 3 Groups, 2 Integrations, and 1 Exposures, in accordance with STScI Exposure Time Calculator tools. This gives a total exposure time of 322.1~s for each pointing and a S/N ratio of 3.1 and 3.2 in F200W and F444W, for targets of 27.4 mag (AB) and 27.5 mag (AB), and a S/N ratio of 5.0 in F200W and F444W, for targets of 26.9 mag (AB) and 27.0 mag (AB), respectively. For efficient telescope steering, we foresee observations grouped into a $9 \times 5$ rectangular mosaic. Full dithering is important in rejecting spurious signals but carries a prohibitively expensive overhead for a wide field survey. However, in our simulations of {\em JWST} observations, we found that sub-pixel dithering with 2-POINT-MEDIUM-WITHNIRISS is sufficient for rejecting spurious signals due to hot pixels and cosmic rays. For a fully  successful program, one should expect to revisit the same field multiple times, which will help build up more robust templates and aid in rejecting cosmic rays.

\subsection{Why cluster lensing does not help find more transients}
%{\it Written by Peter Nugent}

As pointed out in \cite{Sullivan2000} there are three important effects one has to consider in understanding any gain (or loss) in the SN detection rate via a lensing cluster. These are: the area-weighted lensing magnification, the SN rate as a function of redshift and the limiting magnitude of the survey. As can be seen in their Figure~3 and Table~1 for a simulated HST survey, modest gains can be had for $1.5 < z < 2.0$ due to the fact that the intrinsic rates are flat or slowly declining at these redshifts and amplification brings SNe at the lower end of the luminosity function into the realm of detection. Interestingly, the SN detection efficiency {\em drops} for $z < 1.5$ as the area surveyed behind a cluster shrinks in proportion to the amplification factor. Since HST can easily discover SNe without magnification below these redshifts, by reducing the survey area one suffers a net loss. 

This fact is exacerbated in any survey by {\em JWST} to depths of AB~$ = 27$~mag. Since the SN rate starts dropping dramatically beyond a redshift of 2--3 \citep[driven by the $1+z$ time dilation factor; see Figure~2 of][]{Sullivan2000} and {\em JWST} is already sensitive to even the lowest end of the luminosity function for Type~Ia SNe at these redshifts (and more than 70\% of the CC~SNe), a lensing cluster search provides no net benefit. In fact, there would be appreciable loss due to the drop in effective area.

\subsection{Probabilistic Target Classification}

\begin{figure}[H]%[ht]
% \begin{center}
\centering
\includegraphics[width=0.7\linewidth]{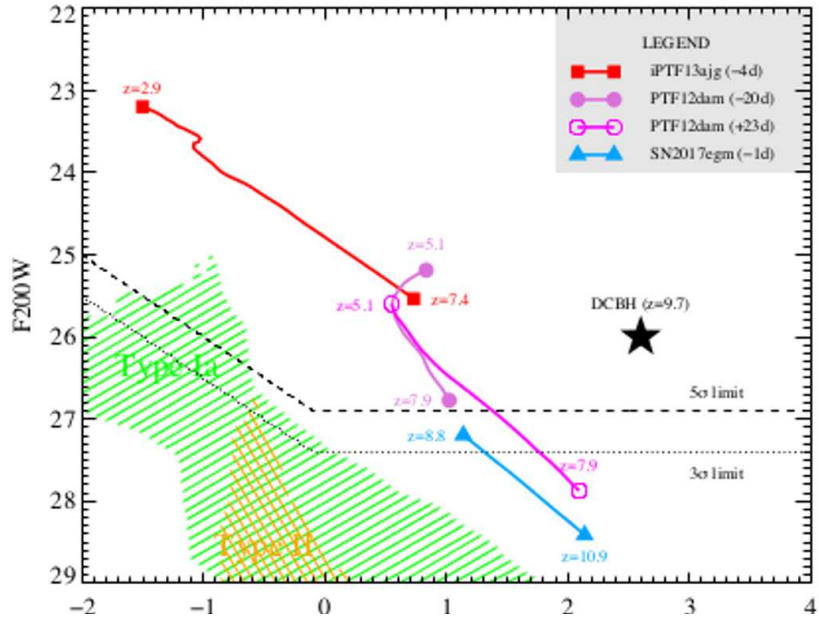}
% \end{center}
%\caption{Is the resolution any better? If not, the original is needed from Lifan.}
%\label{fig:newcmag}
%\end{figure}
\caption{The color--magnitude diagram of Type~Ia SN, SLSNe, and DCBH at various redshifts. The numbers show the redshifts of the targets.}
\label{fig:CMAG}
\end{figure}

With our choice of filters, we are able to make a preliminary screening by the variability of the sources and their location on the color--magnitude diagram. Figure~\ref{fig:CMAG} shows as an example how different objects may be classified. It is clear from the figure that DCBH stand out clearly as extremely red objects, whereas SLSNe and Type~Ia SNe fall on distinct areas on the color--magnitude diagram.  

%\begin{figure}[H]
%\centering
%\includegraphics[width=0.6\textwidth]{FigCMAG.eps}

Ambiguities are inevitable and more data will be needed. We plan to acquire deep images in grizY bands from the ground to supplement the {\em JWST} data. This will be helpful in eliminating brown dwarfs, whose lack of variability at K band \citep{Khandrika2013} escapes triggering our survey. We remark that at the redshift of interest, SLSNe are most likely to be much brighter than their host galaxies and may actually appear ``hostless". This applies also to DCBHs.

\subsubsection{Brown dwarfs}

These objects are redder than brown dwarfs, as can be seen from Figures~18 and 19 calculated by Simpson and Baron using models of Barman et al. (in preparation).

\begin{figure}[H]%[ht]
\centering
\includegraphics[angle=-0,width=0.7\textwidth]{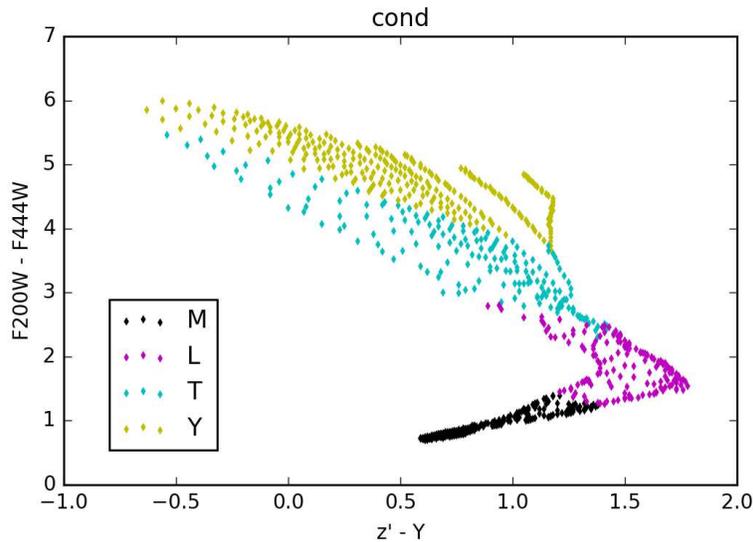}
\vspace{-2 truemm}
\caption{Colors of brown dwarfs of types M, L, T, Y.}%What does cond mean?
\label{fig18}
\end{figure}

\begin{figure}[H]%[h]
\centering
\includegraphics[angle=-0,width=0.7\textwidth]{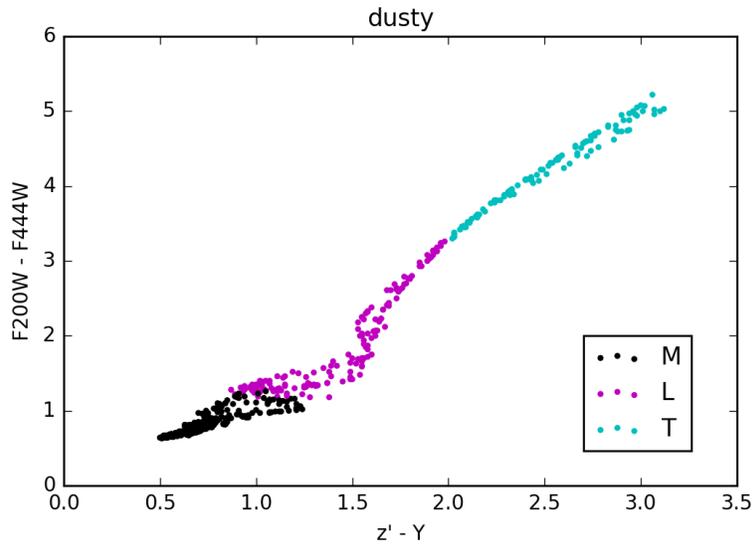}
\vspace{-2 truemm}
\caption{Colors of brown dwarfs with dusty atmospheres.}
\label{fig19}
\end{figure}

% ==============================================================================
\section{Facilitating First Transients discovery}
\label{section:FFTD}

The vision of the {\em JWST} is to discover the first stars to appear in the Universe. For this to be accomplished in fact with this facility we must see not their faint birth but their bright SNe. The FLARE project %{\it First Transients} ERS 
would lay the foundation to do this over the life of the {\em JWST} mission %. It could do this 
by creating a James Webb Transient Factory (JWTF), in which all the mission's repeat imaging data would be analysed for transients.

\subsection{The design of the FLARE project}

\subsubsection{Observations and deliverables with {\em JWST}}
Our %ERS 
observing plan is tailored to deliver some transients from the EoR over the first six months, both single supernovae and transients resulting from the formation by redshift 6 of SMBHs. %Supermassive Black Holes.
But according to \S\ref{section:TSSG} the highest redshift supernovae are rare enough that only operation of a JWTF for five years will find them.
%{\color{red}This section needs to be supplemented by Avishay Gal-Yam.}

%Our ERS 
Appropriate deliverables will therefore include the software for a JWTF, developed at the Weizmann and the Mitchell Institutes, which could  be operated at Space Telescope Science Institute, and also, through a collaboration with Swinburne University and drawing on the Millennium Simulation and other similar databases, a module of TAO which will allow full simulation of all GO proposals for imaging with {\em JWST} by the proposers themselves in the course of preparing Cycle 2 and later cycles.

TAO houses data from popular cosmological $N$-body simulations and galaxy formation models, primarily focused on survey science. Mock catalogues can be built from the database without the need for any coding. Results can be funneled through higher-level modules to generate SEDs, build custom light-cones and images. %TAO is accessible from anywhere you can %access the internet.

TAO will be expanded to include more detailed modelling of the galaxies and SNe at high redshift relevant to the {\em JWST}, and new tools to produce mock observations that mimic those that JWTF will process. This will allow predictions to be made using the advanced simulations that TAO hosts and interpretation of the results as they arrive.

\subsubsection{Supporting observations and followup strategies}
To realise its full potential, time domain astronomy makes serious demands on cadence, followup, and multiwavelength coverage. High-redshift investigations are less demanding than local ones due to time dilation, and we anticipate that NEP revisits can be proposed with normal STScI planned annual cycles. Spectroscopic followup of targets of opportunity with NIRSpec and ELT instruments may also be a modest imposition on these facilities. Deep ground based reference fields at shorter wavelengths, however, should be initiated immediately for the purpose of eliminating foreground objects. The Subaru Stategic Program \citep{Aihara2017} is a model for such data and adding the NEP to the current set of fields seems to us to be a priority. X-ray followup of {\em JWST} high-redshift transients will also be vital.

% ==============================================================================
\section{Conclusion}
\label{section:Conclusion}

The definitive image of the unmistakable structure of the early Universe is the $WMAP$ and $Planck$ iconic picture \citep{Bennett2003,Ade2016}. The vision of {\em JWST} is to link this structure and the associated precision cosmology to the first stars formed from the pristine gas out there in protogalactic clumps. Those first stars are individually too faint, and even first galaxies will still be challenging for {\em JWST}, especially if a representative sample be studied.  

In this white paper, we have demonstrated that instead of hammering at still images, observations of transients offer an elegant alternative method of characterizing the early Universe. Information about stellar populations and the IMF is encapsulated in SLSNe, and TDEs (and the intrinsic variability of the accretion flow) can trace the mysterious build-up of billion solar mass SMBHs within few hundreds of millions of years. Detecting and characterizing first SLSNe and TDEs in sufficient numbers are the goals of the FLARE project for {\em JWST}.  

We have shown that monitoring with {\em JWST} a 0.1~square degree field at 2 and 4~$\mu$m and moderate depth can achieve these goals. Ground based deep imaging at shorter wavelengths will complete the inventory of the field and help avoid false alarms.  

After this quantitative confirmation at the level of transient categories, we will move on to numerical simulations to determine similarly quantitatively how many events need to be found and which combinations of time baseline, cadence, colour information, and field size yield the highest success rates in the classification of high-redshift transients. This will deliver a solid justification of the significant investment of telescope time over the life of the mission.

We intend to make immediately public all transients found by the FLARE project. This will enable the community to define and execute follow-up observations of the transients themselves as well as of their environments of which the high-redshift transients represent the tips of their luminosity functions in transparent regions of the early Universe. The $(1+z)$-fold time dilation makes the planning of such efforts well feasible. We will develop and offer tools to perform analogous searches for transients in all other {\em JWST} fields which throughout the lifetime of {\em JWST} require repeated observations. This will lead to an enlarged homogeneous real-time database of high-redshift transients at no extra cost.  
%The definitive image of the early universe structured unmistakably  as we know it today is the WMAP and Planck iconic picture (Bennett et al 2003; Ade et al 2016). The vision of {\em JWST} is to link that to the first stars made from the pristine gas seen there in protogalactic clumps. As those truly first stars  are individually too faint, the goal of the FLARE project is to `save the day' and go after the first SLSNe, which are expected to be visible, instead. Secondly, there is the mystery of how billion solar mass SMBHs are built in hundreds of millions of years. In FLARE we are going after those too.

%In this white paper we have made the case that a 0.1 square degree transient survey at 2 and 4 microns, together with ground based supporting deep imaging, can realise this goal. We intend to move on from the proof of principle in this paper to detailed simulations, justifying the considerable investment of telescope time in annual cycles over the life of the mission.

\bibliographystyle{aasjournal}
\bibliography{newbib.bib}

% ==============================================================================
\appendix

\section*{Appendix: Pop~III SNe}
\begin{figure}[H]%[ht]
\centering
%\begin{tabular}{cc}
\includegraphics[width=0.45\linewidth]{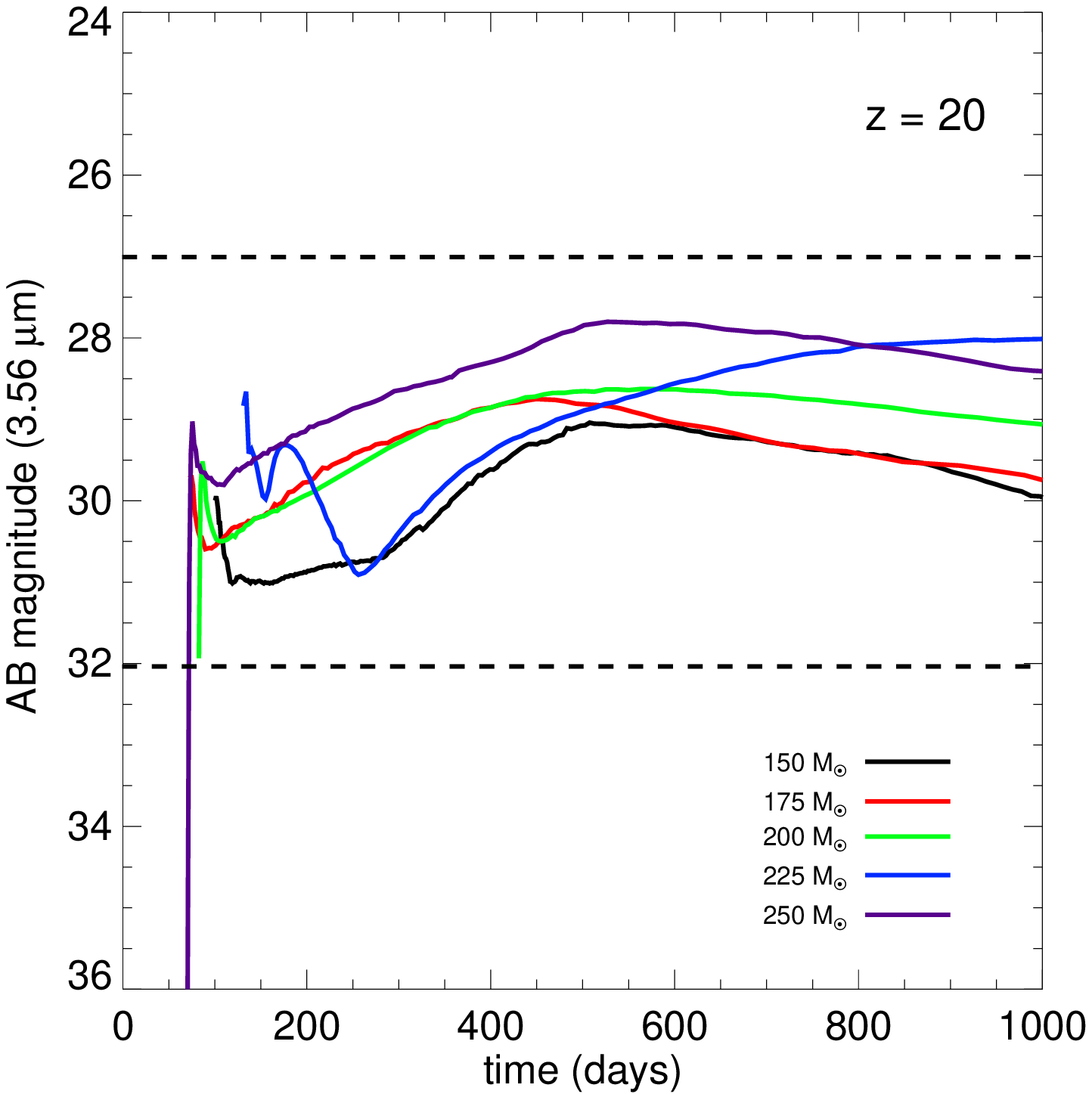}
\includegraphics[width=0.45\linewidth]{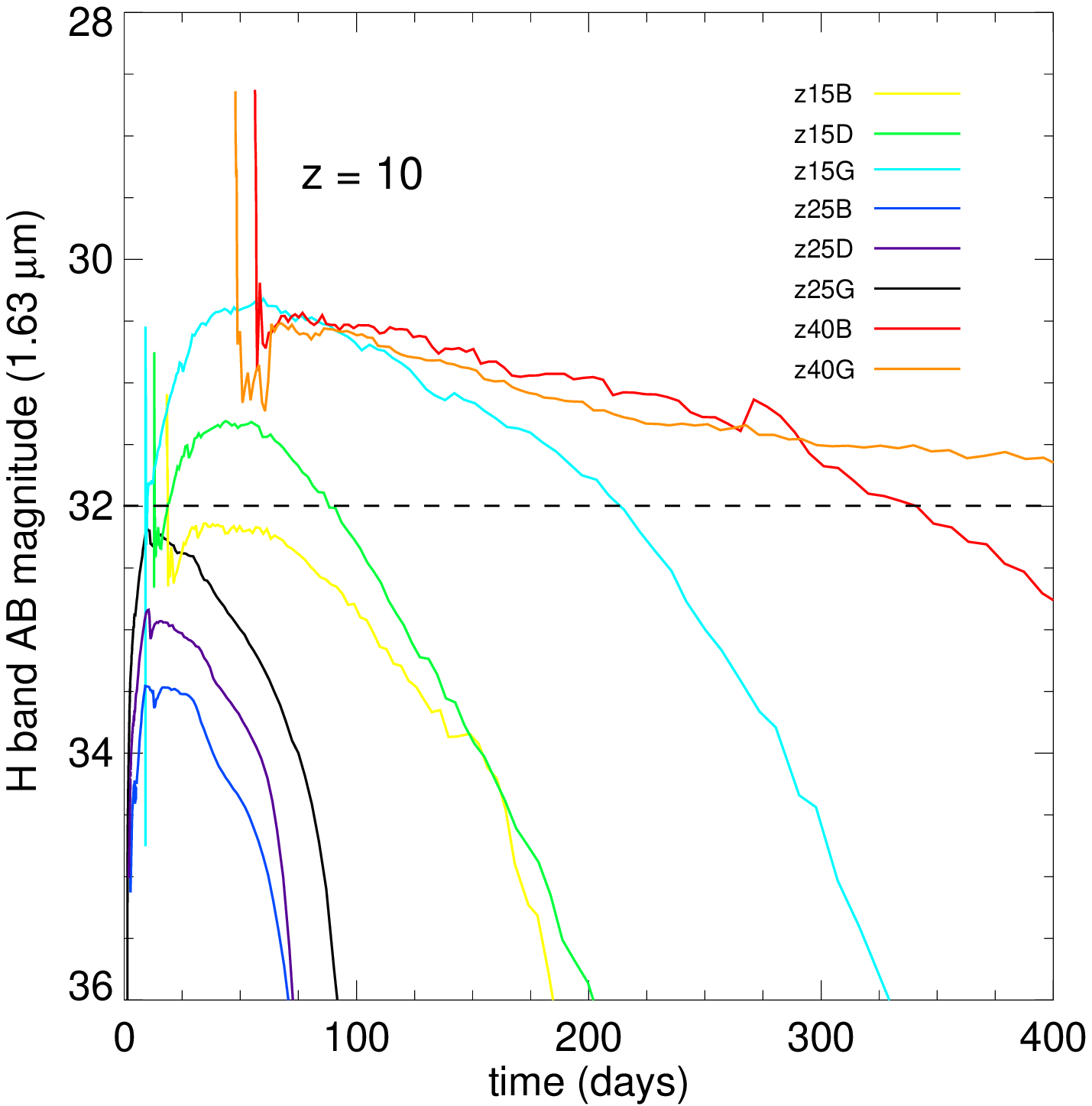}
%\end{tabular}
%\end{center}
\caption{Left panel: light curves for 150, 175, 200, 225 and 250~M$_{\odot}$ Pop~III PI~SNe at $z = 20$ at 3.56~$\mu$m. Right panel: light curves for 15, 25, and 40~M$_{\odot}$ CC~SNe at $z = 10$ at 1.63~$\mu$m. Here, the explosion energies are 0.6~foe (B series), 1.2~foe (D series), and 2.4~foe (G series), where 1~foe $= 10^{51}$~erg.}
\label{fig:NIRLCs}
\end{figure}

An extensive campaign of radiation hydrodynamical simulations has shown that PI~SNe, PPI~SNe and Type~IIn SNe will be visible to {\em JWST} and the extremely large telescopes (ELTs) at $z \gapp 20$ \citep{wet12a,wet12b,wet12e,wet13d}. CC~SNe will be visible to these telescopes at $z = 10$--20 \citep{wet12c} and rotating PI~SNe and hypernovae will be visible out to $z \sim 10$ \citep{smidt13a,smidt14a}. NIR light curves for 150--250~M$_{\odot}$ PI~SNe and 15--40~M$_{\odot}$ CC~SNe are shown in Figure~\ref{fig:NIRLCs}. These studies demonstrate that the optimum wavelengths for observing SNe at $z = 10$--20 are 2--5~$\mu$m. {\em JWST} is uniquely qualified to detect these events because its 40~K temperatures and low thermal noise vastly simplify its systematics in comparison to ground-based ELTs, which must contend with much greater instrument noise and atmospheric transmission in the NIR.

Detections of both PI and CC~SN, whose light curves are easily distinguishable, would enable a rough IMF to be built up over time as more events are found and binned by progenitor mass.  Since both types of explosion are needed to do this, {\em JWST} will only be able to probe the IMF of early stars up to $z \sim 15$ unless gravitational lensing reveals CC~SNe at higher redshifts \citep{pan12a,wet13c}. The possibility that lensed CC~SNe could extend constraints on the Pop~III IMF up to $z \sim 20$, the era of first light, may warrant the inclusion of a number of galaxy cluster lenses with large Einstein radii in the proposed %ERS 
field. Twenty five such clusters have now been studied in the CLASH and Frontier Fields programs by {\em HST} \citep{clash}. Absent such detections, the discovery of PI~SNe alone by {\em JWST} above $z \sim 15$ would still determine if primordial star formation led to stellar masses in excess of $\sim 100$~M$_{\odot}$ and place upper limits on global SFRs at this epoch.

\begin{figure}[H]%[ht]
\centering
% \begin{center}
%\begin{tabular}{cc}
\includegraphics[width=0.65\linewidth]{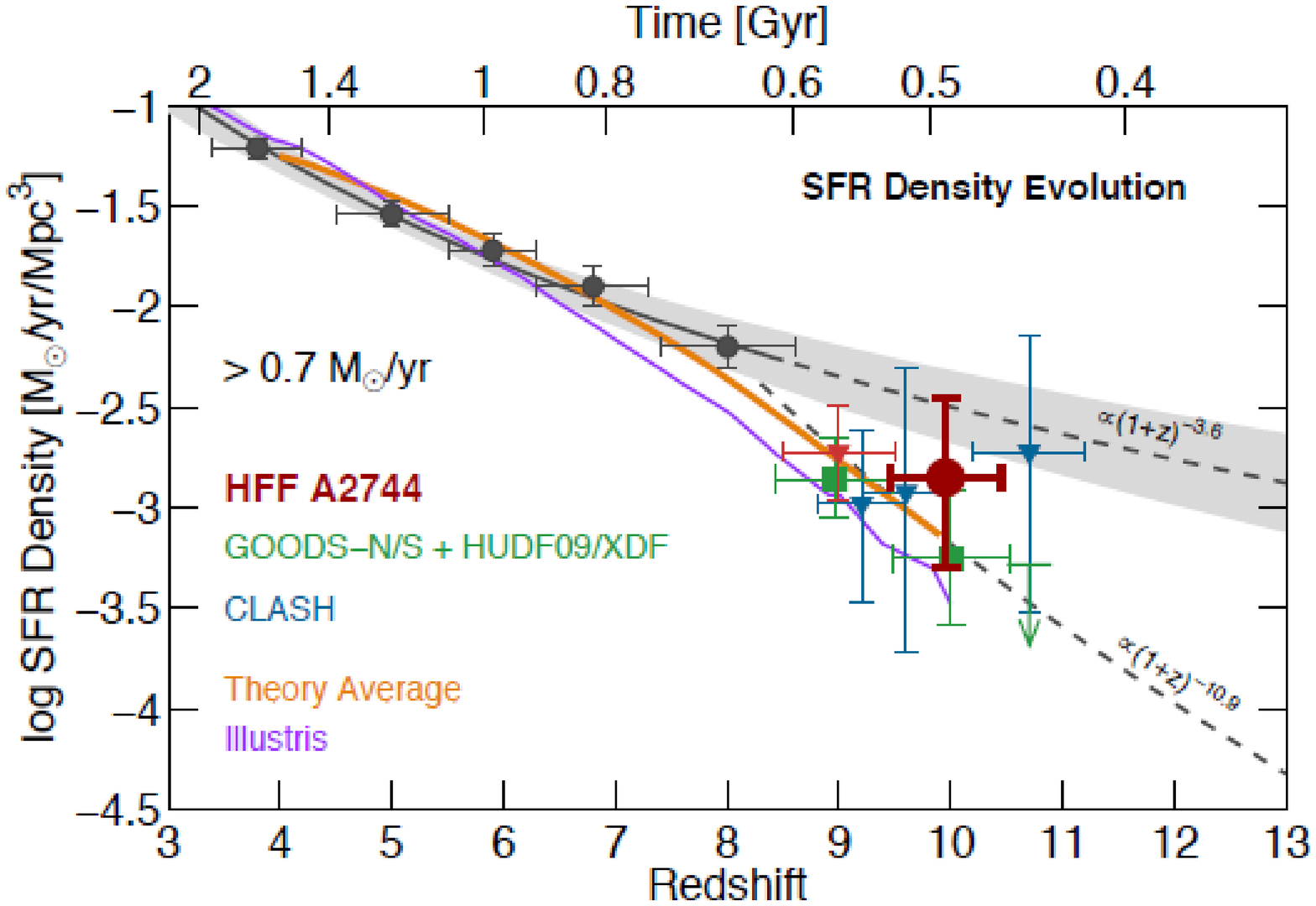}%f4.eps}%SFRs}
\includegraphics[width=0.4\linewidth]{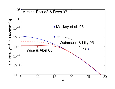}
%\end{tabular}
% \end{center}
\caption{Global SFRs as a function of redshift \citep[from][]{Oesch2015}.
Two possible decline rates are shown. %The purple and green 
%bands are upper and lower limits from GRBs and low-luminosity high-redshift galaxies, 
%respectively \citep{re12}. The lines are rates taken from the simulations listed in the
%legend.  
Right panel: PI~SN rates in number per year per {\em JWST} field of view above a given redshift \citep{hum12}. There are baryonic feedback mechanisms within the galactic halos where these massive stars form. We show the upper limit for weak feedback (blue line), strong feedback (red line), and an intermediate case (dashed red line). The symbols denote additional estimates from the literature.}
\label{fig:SFRs}
\end{figure}

The main challenge to finding high-redshift transients is their low surface densities, which require large fields. Since cosmic SFRs inferred from observations of GRBs and low-luminosity, high-redshift galaxies have only been extrapolated up to $z = 15$ \citep{re12}, one must resort to numerical simulations for estimates of SFRs at earlier times. As shown in the left panel of Figure~\ref{fig:SFRs}, these rates vary by a factor of 300 at the highest redshifts, highlighting the present uncertainties in them. However, even at $z \sim 20$--25, {\em JWST} is predicted to find a number of PI~SNe \citep[right panel of Figure~\ref{fig:SFRs};][]{hum12}. The optimal search strategy is a mosaic approach with only modestly deep individual exposures. Our proposed %ERS 
field is well suited to push into this wide-field regime to place constraints on the SFR density of massive Pop~III stars.  

However, even the unlikely failure of {\em JWST} to harvest any transients in our FLARE %ERS
field at $z \gapp 15$ would place useful, strong upper limits on global SFRs in the primordial Universe and could even shed light on the properties of dark matter. Models of early structure formation with warm dark matter predict a suppression of power on small scales that delays Pop~III star formation to $z \sim 15$, so the failure to detect SNe at earlier times might be a signature of such suppression \citep{magg16}.  Furthermore, any detections of SNe at $z \sim 10$--20 would still trace the rise of stellar populations in the first galaxies and put global SFRs on much firmer footing, even though they might not all be Pop~III events due to chemical enrichment by the first SNe.

\end{document}